\documentclass{paper}

\usepackage{multirow}
\usepackage{longtable}
\usepackage{url}
\usepackage{booktabs}
\usepackage{times}
\usepackage{amsthm}
\usepackage{amssymb}
\usepackage{multicol}
\usepackage{amsmath}
\usepackage{graphicx}
\usepackage{array}

\newcommand{\heuristic}[1]{{\small\textsf{{#1}}}}
\newcommand{\ifamily}[1]{{\small\textsf{{#1}}}}

\newcommand{\OptPlain}[1]{\mbox{#1-opt}}
\newcommand{\Opt}[1]{\heuristic{\OptPlain{#1}}}

\newcommand{\TwooptPlain}{\OptPlain{2}}
\newcommand{\Twoopt}{\Opt{2}}

\newcommand{\ThreeoptPlain}{\OptPlain{3}}
\newcommand{\Threeopt}{\Opt{3}}

\newcommand{\KoptPlain}{\OptPlain{$k$}}
\newcommand{\Kopt}{\Opt{$k$}}

\newcommand{\OptStarPlain}[1]{\mbox{#1-opt*}}
\newcommand{\OptStar}[1]{\heuristic{\OptStarPlain{#1}}}

\newcommand{\VoptPlain}{\OptPlain{v}}
\newcommand{\Vopt}{\Opt{v}}

\newcommand{\OneDVPlain}{1DV}
\newcommand{\OneDV}{{\heuristic{\OneDVPlain}}}

\newcommand{\TwoDVPlain}{2DV}
\newcommand{\TwoDV}{{\heuristic{\TwoDVPlain}}}

\newcommand{\MDVPlain}{$s$DV}
\newcommand{\MDV}{{\heuristic{\MDVPlain}}}

\newcommand{\VDI}{VDI}

\newcommand{\CombPlain}[2]{#1$_\text{#2}$}
\newcommand{\Comb}[2]{\heuristic{\CombPlain{#1}{#2}}}

\newcommand{\OneDVtwoPlain}{\CombPlain{\OneDVPlain}{2}}
\newcommand{\OneDVthreePlain}{\CombPlain{\OneDVPlain}{3}}
\newcommand{\OneDVVPlain}{\CombPlain{\OneDVPlain}{\text{v}}}

\newcommand{\TwoDVtwoPlain}{\CombPlain{\TwoDVPlain}{2}}
\newcommand{\TwoDVthreePlain}{\CombPlain{\TwoDVPlain}{3}}
\newcommand{\TwoDVVPlain}{\CombPlain{\TwoDVPlain}{\text{v}}}

\newcommand{\MDVtwoPlain}{\CombPlain{\MDVPlain}{2}}
\newcommand{\MDVthreePlain}{\CombPlain{\MDVPlain}{3}}
\newcommand{\MDVVPlain}{\CombPlain{\MDVPlain}{\text{v}}}

\newcommand{\OneDVtwo}{\heuristic{\OneDVtwoPlain}}
\newcommand{\OneDVthree}{\heuristic{\OneDVthreePlain}}
\newcommand{\OneDVV}{\heuristic{\OneDVVPlain}}

\newcommand{\TwoDVtwo}{\heuristic{\TwoDVtwoPlain}}
\newcommand{\TwoDVthree}{\heuristic{\TwoDVthreePlain}}
\newcommand{\TwoDVV}{\heuristic{\TwoDVVPlain}}

\newcommand{\MDVtwo}{\heuristic{\MDVtwoPlain}}
\newcommand{\MDVthree}{\heuristic{\MDVthreePlain}}
\newcommand{\MDVV}{\heuristic{\MDVVPlain}}

\newcommand{\Trivial}{{\heuristic{Trivial}}}
\newcommand{\Greedy}{{\heuristic{Greedy}}}
\newcommand{\MaxRegret}{{\heuristic{Max-Regret}}}
\newcommand{\ROM}{{\heuristic{ROM}}}

\newcommand{\Chain}{{\heuristic{Chain}}}
\newcommand{\Multichain}{{\heuristic{Multichain}}}

\newcommand{\random}{{\ifamily{Random}}}
\newcommand{\clique}{{\ifamily{Clique}}}
\newcommand{\geometric}{{\ifamily{Geometric}}}
\newcommand{\GP}{{\ifamily{GP}}}
\newcommand{\squareroot}{{\ifamily{SquareRoot}}}
\newcommand{\product}{{\ifamily{Product}}}

\newtheorem{theorem}{Theorem}

\DeclareMathOperator*{\argmin}{argmin}

\begin{document}

\title{Local Search Heuristics for the Multidimensional Assignment Problem\thanks{A preliminary version of this paper was published in Golumbic Festschrift, volume 5420 of Lect. Notes Comput. Sci., pages 100--115, Springer, 2009.}}

\author{Gregory Gutin\thanks{Royal Holloway, University of London, \texttt{gutin@cs.rhul.ac.uk}} \and Daniel Karapetyan\thanks{Royal Holloway, University of London, \texttt{daniel.karapetyan@gmail.com}}}


\date{}

\maketitle

\begin{abstract}
The Multidimensional Assignment Problem (MAP) (abbreviated $s$-AP in the case of $s$ dimensions) is an extension of the well-known assignment problem.  The most studied case of MAP is 3-AP, though the problems with larger values of $s$ also have a large number of applications.  We consider several known neighborhoods, generalize them and propose some new ones.  The heuristics are evaluated both theoretically and experimentally and dominating algorithms are selected.  We also demonstrate a combination of two neighborhoods may yield a heuristics which is superior to both of its components.

\end{abstract}

\section{Introduction}
\label{sec:introduction}

The \emph{Multidimensional Assignment Problem} (MAP) (abbreviated $s$-AP in the case of $s$ dimensions, also called \emph{(axial) Multi Index Assignment Problem}, MIAP,~\cite{Bandelt2004,Pardalos2000}) is a well-known optimization problem.  It is an extension of the \emph{Assignment Problem} (AP), which is exactly the two dimensional case of MAP\@.  While AP can be solved in the polynomial time~\cite{Kuhn1955}, $s$-AP for every $s \ge 3$ is NP-hard~\cite{Garey1979} and inapproximable~\cite{Burkard1996}\footnote{Burkard et al.\ show it for a special case of 3-AP and since 3-AP is a special case of $s$-AP the result can be extended to the general MAP}\@.  The most studied case of MAP is the case of three dimensions~\cite{Aiex2005,Andrijich2001,Balas1991,Crama1992,Huang2006,Spieksma2000} though the problem has a host of applications for higher numbers of dimensions, e.g., in matching information from several sensors (data association problem), which arises in plane tracking~\cite{Murphey1998,Pardalos2000b}, computer vision~\cite{Veenman2003} and some others~\cite{Andrijich2001,Bandelt2004,Burkard1999}, in routing in meshes~\cite{Bandelt2004}, tracking elementary particles~\cite{Pusztaszeri1996}, solving systems of polynomial equations~\cite{Bekker2005}, image recognition~\cite{Grundel2004}, resource allocation~\cite{Grundel2004}, etc.

For a fixed $s \ge 2$, the problem $s$-AP is stated as follows.  Let $X_1 = X_2 = \ldots = X_s = \{ 1, 2, \ldots, n \}$.  We will consider only vectors that belong to the Cartesian product $X = X_1 \times X_2 \times \ldots \times X_s$.  Each vector $e \in X$ is assigned a non-negative weight $w(e)$.  For a vector $e \in X$, the component $e_j$ denotes its $j$th coordinate, i.e., $e_j \in X_j$.  A collection $A$ of $t \le n$ vectors $A^1, A^2, \ldots, A^t$ is a \emph{(feasible) partial assignment} if $A^i_j \neq A^k_j$ holds for each $i \neq k$ and $j \in \{ 1, 2, \ldots, s \}$.  The \emph{weight} of a partial assignment $A$ is $w(A) = \sum_{i=1}^t w(A^i)$.  An \emph{assignment} (or \emph{full assignment}) is a partial assignment with $n$ vectors.  The objective of $s$-AP is to find an assignment of minimal weight.  

We also provide a \emph{permutation form} of the assignment which is sometimes more convenient.  Let $\pi_1, \pi_2, \ldots, \pi_s$ be permutations of $X_1, X_2, \ldots, X_s$ respectively.  Then $\pi_1 \pi_2 \ldots \pi_s$ is an assignment with the weight $\sum_{i=1}^n w(\pi_1(i) \pi_2(i) \ldots \pi_s(i))$.

It is obvious that one permutation, say the first one, may be fixed without any loss of generality: $\pi_1 = 1_n$, where $1_n$ is the identity permutation of size $n$.  Then the objective of the problem is as follows:
$$
\min_{\pi_2, \ldots, \pi_s} \sum_{i=1}^n w(i \pi_1(i) \ldots \pi_s(i)) \ .
$$

A graph formulation of the problem is as follows.  Having an $s$-partite graph $G$ with parts $X_1$, $X_2$, \ldots, $X_s$, where $|X_i| = n$, find a set of $n$ disjoint cliques in $G$ of the minimal total weight if every clique $e$ in $G$ is assigned a weight $w(e)$.

Finally, an integer programming formulation of the problem is as follows.
$$
\min \sum_{i_1 \in X_1, \ldots, i_s \in X_s} w(i_1 \ldots i_s) \cdot x_{i_1 \ldots i_s}
$$
subject to
$$
\sum_{i_2 \in X_2, \ldots, i_s \in X_s} x_{i_1 \ldots i_s} = 1 \qquad {\forall i_1 \in X_1} \mbox{ ,}
$$
$$
\ldots
$$
$$
\sum_{i_1 \in X_1, \ldots, i_{s-1} \in X_{s-1}} x_{i_1 \ldots i_s} = 1 \qquad {\forall i_s \in X_s} \mbox{ ,}
$$
where $x_{i_1 \ldots i_s} \in \{ 0, 1 \}$ for all $i_1$, \ldots, $i_s$ and $|X_1| = \ldots = |X_s| = n$.

Sometimes the problem is formulated in a more general way if $|X_1| = n_1$, $|X_2| = n_2$, \ldots, $|X_s| = n_s$ and the requirement $n_1 = n_2 = \ldots = n_s$ is omitted.  However this case can be easily transformed into the problem described above by complementing the weight matrix to an $n \times n \times \ldots \times n$ matrix with zeros, where $n = \max_i n_i$.

The problem was studied by many researchers.  Several special cases of the problem were intensively studied in the literature (see~\cite{Kuroki2007} and references there) and for few classes of instances polynomial time exact algorithms were found, see, e.g.,~\cite{Burkard1996a,Burkard1996,Isler2005}.  In many cases MAP remains hard to solve~\cite{Kuroki2007,Spieksma1996}.  For example, if there are three sets of points of size $n$ on a Euclidean plain and the objective is to find $n$ triples of points, one from each set, such that the total circumference or area of the corresponding triangles is minimal, the corresponding 3-AP is still NP-hard~\cite{Spieksma1996}.  The asymptotic properties of some special instance families are studied in~\cite{Grundel2004}.  

As regards the solution methods, apart from exact and approximation algorithms~\cite{Balas1991,Crama1992,Kuroki2007,Pasiliao2005,Pierskalla1968}, several heuristics including construction heuristics~\cite{Balas1991,Gutin2008,GK_MAP_Construction,Oliveira2004}, greedy randomized adaptive search procedures~\cite{Aiex2005,Murphey1998,Oliveira2004,Robertson2001}, metaheuristics~\cite{Clemons2004,Huang2006} and parallel heuristics~\cite{Oliveira2004} are presented in the literature.  Several local search procedures are proposed and discussed in~\cite{Aiex2005,Balas1991,Bandelt2004,Burkard1996,Clemons2004,Huang2006,Oliveira2004,Robertson2001}.

The difference between the construction heuristics and local search is sometimes crucial.  While a construction heuristic generates a solution from scratch and, thus, has some solution quality limitation, local search is intended to improve an existing solution and, thus, can be used after a construction heuristic or as a part of a more sophisticated heuristic, so called metaheuristic.

The contribution of our paper is in collecting and generalizing all local search heuristics known from the literature, proposing some new ones and detailed theoretical and evaluating them both theoretically and experimentally.  For the purpose of experimental evaluation we also thoroughly discuss, classify the existing instance families and propose some new ones.

In this paper we consider only the general case of MAP and, thus, all the heuristics which rely on the special structures of the weight matrix are not included in the comparison.  We also assume that the number of dimensions $s$ is a small fixed constant while the size $n$ can be arbitrary large.

\section{Heuristics}
\label{sec:heuristics}

In this section we discuss some well known and some new MAP local search heuristics as well as their combinations.

\subsection{Dimensionwise Variations Heuristics}
\label{sec:dv}

The heuristics of this group were first introduced by Bandelt et al.~\cite{Bandelt2004} for MAP with decomposable costs.  However, having a very large neighborhood (see below), they are very efficient even in the general case.  The fact that this approach was also used by Huang and Lim as a local search procedure for their memetic algorithm~\cite{Huang2006} confirms its efficiency.

The idea of the dimensionwise variation heuristics is as follows.  Consider the initial assignment $A$ in the permutation form $A = \pi_1 \pi_2 \ldots \pi_s$ (see Section~\ref{sec:introduction}).  Let $p(A, \rho_1, \rho_2, \ldots, \rho_s)$ be an assignment obtained from $A$ by applying the permutations $\rho_1, \rho_2, \ldots, \rho_s$ to $\pi_1, \pi_2, \ldots, \pi_s$ respectively:
\begin{equation}
\label{eq:p}
p(A, \rho_1, \rho_2, \ldots, \rho_s) = \rho_1(\pi_1) \rho_2(\pi_2) \ldots \rho_s(\pi_s) \ .
\end{equation}
Let $p_D(A, \rho)$ be an assignment $p(A, \rho_1, \rho_2, \ldots, \rho_s)$, where $\rho_j = \rho$ if $j \in D$ and $\rho_j = 1_n$ otherwise ($1_n$ is the identity permutation of size $n$): 
\begin{equation}
\label{eq:p_D}
p_D(A, \rho) = p\left(A, \left\{ \begin{array}{@{}l@{~}l@{}}\rho & \text{if } 1 \in D\\ 1_n & \text{otherwise}\end{array} \right., \left\{ \begin{array}{@{}l@{~}l@{}}\rho & \text{if } 2 \in D\\ 1_n & \text{otherwise}\end{array} \right., \ldots, \left\{ \begin{array}{@{}l@{~}l@{}}\rho & \text{if } s \in D\\ 1_n & \text{otherwise}\end{array} \right.\right) \ .
\end{equation}
On every iteration, the heuristic selects some nonempty set $D \subsetneq \{ 1, 2, \ldots, s \}$ of dimensions and searches for a permutation $\rho$ such that $w(p_D(A, \rho))$ is minimized.

For every subset of dimensions $D$, there are $n!$ different permutations $\rho$ but the optimal one can be found in the polynomial time.  Let $swap(u, v, D)$ be a vector which is equal to vector $u$ in all dimensions $j \in \{ 1, 2, \ldots, s \} \setminus D$ and equal to vector $v$ in all dimensions $j \in D$:
\begin{equation}
\label{eq:swap}
swap(u, v, D)_j = \left\{ \begin{array}{l @{\quad} l}
u_j & \text{if } j \notin D \\
v_j & \text{if } j \in D \\
\end{array} \right. \text{\qquad for } j = 1, 2, \ldots, s.
\end{equation}
Let matrix $[M_{i,j}]_{n \times n}$ be constructed as
\begin{equation}
\label{eq:M}
M_{i,j} = w(swap(A^i, A^j, D)) \ .
\end{equation}
It is clear that the solution of the corresponding 2-AP is exactly the required permutation $\rho$.  Indeed, assume there exists some permutation $\rho'$ such that $w(p_D(A, \rho')) < w(p_D(A, \rho))$.  Observe that $p_D(A, \rho) = \{ swap(A^i, A^{\rho(i)}, D) :\ i \in \{ 1, 2, \ldots, n \} \}$.  Then we have
$$
\sum_{i = 1}^n w(swap(A^i, A^{\rho'(i)}, D)) < \sum_{i = 1}^n w(swap(A^i, A^{\rho(i)}, D)) \ .
$$
Since $w(swap(A^i, A^{\rho(i)}, D)) = M_{i, \rho(i)}$, the sum $\sum_{i = 1}^n w(swap(A^i, A^{\rho(i)}, D))$ is already minimized to the optimum and no $\rho'$ can exist.

\bigskip

The neighborhood of a dimensionwise heuristic is as follows:
\begin{equation}
\label{eq:dv_neighborhood}
N_\text{DV}(A) = \big\{ p_D(A, \rho) :\ D \in \mathcal{D} \text{ and $\rho$ is a permutation} \big\} \ ,
\end{equation}
where $\mathcal{D}$ includes all dimension subsets acceptable by a certain heuristic.  Observe that
\begin{equation}
\label{eq:p_D_symmetry}
p_D(A, \rho) = p_{\overline{D}}(A, \rho^{-1}) \ ,
\end{equation}
where $\rho^{-1}(\rho) = \rho(\rho^{-1}) = 1_s$ and $\overline{D} = \{ 1, 2, \ldots, s \} \setminus D$, and, hence,
\begin{equation}
\label{eq:P_D_symmetry}
\big\{ p_D(A, \rho) :\ \text{$\rho$ is a permutation} \big\} = \big\{ p_{\overline{D}}(A, \rho) :\ \text{$\rho$ is a permutation} \big\}
\end{equation}
for any $D$.  From (\ref{eq:P_D_symmetry}) and the obvious fact that $p_\varnothing(A, \rho) = p_{\{ 1, 2, \ldots, s \}}(A, \rho) = A$ for any $\rho$ we introduce the following restrictions for $\mathcal{D}$:
\begin{equation}
\label{eq:D_restrictions}
D \in \mathcal{D} \Rightarrow\overline{D} \notin \mathcal{D} \text{\quad\ and \quad} \varnothing, \{ 1, 2, \ldots, s \} \notin \mathcal{D} \ .
\end{equation}
With these restrictions, one can see that for any pair of distinct sets $D_1, D_2 \in \mathcal{D}$ the equation $p_{D_1}(A, \rho_1) = p_{D_2}(A, \rho_2)$ holds if and only if $\rho_1 = \rho_2 = 1_n$.  Hence, the size of the neighborhood $N_\text{DV}(A)$ is
\begin{equation}
\label{eq:dv_neighborhood_size}
|N_\text{DV}(A)| = |\mathcal{D}| \cdot (n! - 1) + 1 \ .
\end{equation}

In~\cite{Bandelt2004} it is decided that the number of iterations should not be exponential with regards to neither $n$ nor $s$ while the size of the maximum $\mathcal{D}$ is $|\mathcal{D}| = 2^{s-1} - 1$.  Therefore two heuristics, LS1 and LS2, are evaluated in~\cite{Bandelt2004}.  LS1 includes only singleton values of $D$, i.e., $\mathcal{D} = \{ D :\ |D| = 1 \}$; LS2 includes only doubleton values of $D$, i.e., $\mathcal{D} = \{ D :\ |D| = 2 \}$.  It is surprising but according to both~\cite{Bandelt2004} and our computational experience, the heuristic LS2 produces worse solutions than LS1 though it obviously has larger neighborhood and larger running times.  We improve the heuristic by allowing $|D| \le 2$, i.e., $\mathcal{D} = \{ D :\ |D| \le 2 \}$.  This does not change the theoretical time complexity of the algorithm but improves its performance.  The heuristic LS1 is called \OneDV\ in our paper; LS2 with $|D| \le 2$ is called \TwoDV.  We also assume (see Section~\ref{sec:introduction}) that the value of $s$ is a small fixed constant and, thus, introduce a heuristic \MDV\ which enumerates all feasible (recall (\ref{eq:D_restrictions})) $D \subset \{ 1, 2, \ldots, s \}$.

The order in which the heuristics take the values $D \in \mathcal{D}$ in our implementations is as follows.  For \OneDV\ it is $\{ 1 \}$, $\{ 2 \}$, \ldots, $\{ s \}$.  \TwoDV\ begins as \OneDV\ and then takes all pairs of dimensions: $\{ 1, 2 \}$, $\{ 1, 3 \}$, \ldots, $\{ 1, s \}$, $\{ 2, 3 \}$, \ldots, $\{ s - 1, s \}$.  Note that because of~(\ref{eq:D_restrictions}) it enumerates no pairs of vectors for $s = 3$, and for $s = 4$ it only takes the following pairs: $\{ 2, 3 \}$, $\{ 2, 4 \}$ and $\{ 3, 4 \}$.  \MDV\ takes first all sets $D$ of size 1, then all sets $D$ of size 2 and so on up to $|D| = \lfloor s / 2 \rfloor$.  If $s$ is even then we should take only half of the sets $D$ of size $s / 2$ (recall (\ref{eq:P_D_symmetry})); for this purpose we take all the subsets of $D \subset \{ 2, 3, \ldots, s \}$, $|D| = s / 2$ in the similar order as before.


It is obvious that $N_\text{\OneDVPlain}(A) \subseteq N_\text{\TwoDVPlain}(A) \subseteq N_\text{\MDVPlain}(A)$ for any $s$ however for $s = 3$ all the neighborhoods are equal and for $s = 4$ \TwoDV\ and \MDV\ also coincide.

According to (\ref{eq:D_restrictions}) and (\ref{eq:dv_neighborhood_size}), the neighborhood size of \OneDV\ is 
$$
|N_\text{\OneDVPlain}(A)| = s \cdot (n! - 1) + 1 \ ,
$$
of \TwoDV\ is 
$$
|N_\text{\TwoDVPlain}(A)| = \left\{ \begin{array}{ll}
(2^{s-1} - 1) \cdot (n! - 1) + 1 & \text{if } s \in \{ 3, 4 \} \\
\left(\binom{s}{2} + s\right) \cdot (n! - 1) + 1 & \text{if } s \ge 5
\end{array} \right. \ ,
$$
and of \MDV\ is 
$$
|N_\text{\MDVPlain}(A)| = (2^{s-1} - 1) \cdot (n! - 1) + 1 \ .
$$

The time complexity of every run of DV is $O(|\mathcal{D}| \cdot n^3)$ as every 2-AP takes $O(n^3)$ and, hence, the time complexity of \OneDV\ is $O(s \cdot n^3)$, of \TwoDV\ is $O(s^2 \cdot n^3)$ and of MDV is $O(2^{s-1} \cdot n^3)$.

\subsection{$k$-opt}
\label{sec:kopt}

The \Kopt\ heuristic for 3-AP for $k = 2$ and $k = 3$ was first introduced by Balas and Saltzman~\cite{Balas1991} as a \emph{pairwise} and \emph{triple interchange heuristic}.  \Twoopt\ as well as its variations were also discussed in~\cite{Aiex2005,Clemons2004,Murphey1998,Oliveira2004,Pasiliao2005,Robertson2001} and some other papers.  We generalize the heuristic for arbitrary values of $k$ and $s$.

The heuristic proceeds as follows.  For every subset of $k$ vectors taken in the assignment $A$ it removes all these vectors from $A$ and inserts some new $k$ vectors such that the assignment feasibility is preserved and its weight is minimized.  Another definition is as follows: for every set of distinct vectors $e^1, e^2, \ldots, e^k \in A$ let $X'_j = \{ e_j^1, e_j^2, \ldots, e_j^k \}$ for $j = 1, 2, \ldots, k$.  Let $A' = \{ e'^1, e'^2, \ldots, e'^k \}$ be the solution of this $s$-AP of size $k$.  Replace the vectors $e^1, e^2, \ldots, e^k$ in the initial assignment $A$ with $e'^1, e'^2, \ldots, e'^k$.

The time complexity of \Kopt\ is obviously $O\left( \binom{n}{k} \cdot k!^{s-1} \right)$; for $k \ll n$ it can be replaced with $O(n^k \cdot k!^{s-1})$.  It is a natural question if one can use some faster solver on every iteration.  Indeed, according to Section~\ref{sec:introduction} it is possible to solve $s$-AP of size $k$ in $O(k!^{s-2} \cdot k^3)$.  However, it is easy to see that $k!^{s-1} < k!^{s-2} \cdot k^3$ for every $k$ up to 5, i.e., it is better to use the exhaustive search for any reasonable $k$.  One can doubt that the exact algorithm actually takes $k!^{s-2} \cdot k^3$ operations but even for the lower bound $\Omega(k!^{s-2} \cdot k^2)$ the inequality $k!^{s-1} < k!^{s-2} \cdot k^2$ holds for any $k \le 3$, i.e., for all the values of $k$ we actually consider.

Now let us find the neighborhood of the heuristic.  For some set $\mathcal{I}$ and a subset $I \subset \mathcal{I}$ let a permutation $\rho$ of elements in $\mathcal{I}$ be an \emph{$I$-permutation} if $\rho(i) = i$ for every $i \in \mathcal{I} \setminus I$, i.e., if $\rho$ does not move any elements except elements from $I$\@.  Let $E = \{ e^1, e^2, \ldots, e^k \} \subset A$ be a set of $k$ distinct vectors in $A$\@.  For $j = 2, 3, \ldots, s$ let $\rho_j$ be an $E_j$-permutation, where $E_j = \{ e_j^1, e_j^2, \ldots, e_j^k \}$.  Then a set $W(A, E)$ of all assignments which can be obtained from $A$ by swapping coordinates of vectors $E$ can be described as follows:
$$
W(A, E) = \big\{ p(A, 1_n, \rho_2, \rho_3, \ldots, \rho_s) :\ \rho_j \text{ is an $E_j$-permutation for } j = 2, 3, \ldots, s \big\} \ .
$$
Recall that $1_n$ is the identity permutation of size $n$ and $p(A, \rho_1, \rho_2, \ldots, \rho_s)$ is defined by (\ref{eq:p}).



The neighborhood $N_{k\text{-opt}}(A)$ is defined as follows:
\begin{equation}
\label{eq:KoptNeighborhood}
N_{k\text{-opt}}(A) = \bigcup_{E \subset A, |E| = k} W(A, E) \ .
\end{equation}

Let $Y, Z \subset A$ such that $|Y| = |Z| = k$.  Observe that $W(A, Y) \cap W(A, Z)$ is nonempty and apart from the initial assignment $A$ this intersection may contain assignments which are modified only in the common vectors $Y \cap Z$.  To calculate the size of the neighborhood of \Kopt{} let us introduce $W'(A, E)$ as a set of all assignments in $W(A, E)$ such that every vector in $E$ is modified in at least one dimension, where $E \subset A$ is the set of $k$ selected vectors in the assignment $A$:
$$
W'(A, E) = \big\{ A' \in W(A, E) :\ |A \cap A'| = n - k \big\} \ .
$$
Then the neighborhood $N_{k\text{-opt}}(A)$ of \Kopt{} is
\begin{equation}
N_{k\text{-opt}}(A) = \bigcup_{E \subset A, |E| \le k} W'(A, E)
\end{equation}
and since $W(A, Y) \cap W(A, Z) = \varnothing$ if $Y \neq Z$ we have
\begin{equation}
\label{eq:kopt_neighborhood_size}
|N_{k\text{-opt}}(A)| = \sum_{E \subset A, |E| \le k} |W'(A, E)| = \sum_{i = 0}^k \binom{n}{i} N^i \ ,
\end{equation}
where $N^i = |W(A, E)|$ for any $E$ with $|E| = i$.  Observe that
$$
W'(A, E) = W(A, E) \setminus \bigcup_{E' \subsetneq E} W'(A, E')
$$
and $|W(A, E)| = k!^{s-1}$ for $|E| = k$ and, hence,
\begin{equation}
\label{eq:kopt_n_k}
N^k = k!^{s-1} - \sum_{i = 0}^{k-1}\binom{k}{i} N^{i} \ .
\end{equation}
It is obvious that $N^0 = 1$ since one can obtain exactly one assignment (the given one) by changing no vectors.  From this and~(\ref{eq:kopt_n_k}) we have $N^1 = 0$, $N^2 = 2^{s-1} - 1$ and $N^3 = 6^{s-1} - 3 \cdot 2^{s-1} + 2$.  From this and~(\ref{eq:kopt_neighborhood_size}) follows
\begin{equation}
\label{eq:twoopt_neighborhood_size}
|N_\text{2-opt}(A)| = 1 + \binom{n}{2} (2^{s-1} - 1) \ ,
\end{equation}
\begin{equation}
|N_\text{3-opt}(A)| = 1 + \binom{n}{2} (2^{s-1} - 1) + \binom{n}{3} (6^{s-1} - 3 \cdot 2^{s-1} + 2) \ .
\end{equation}

\bigskip

In our implementation, we skip an iteration if the corresponding set of vectors $E$ either consists of the vectors of the minimal weight ($w(e) = \min_{e \in X} w(e)$ for every $e \in E$) or all these vectors have remained unchanged during the previous run of $k$-opt.

\bigskip

It is assumed in the literature~\cite{Balas1991,Pasiliao2005,Robertson2001} that \Kopt\ for $k > 2$ is too slow to be applied in practice.  However, the neighborhood $N_{k\text{-opt}}$ do not only includes the neighborhood $N_{(k-1)\text{-opt}}$ but also grows exponentially with the growth of $k$ and, thus, becomes very powerful.  We decided to include \Twoopt\ and \Threeopt\ in our research.  Greater values of $k$ are not considered in this paper because of nonpractical time complexity (observe that the time complexity of \Opt{4} is $O(n^4 \cdot 24^{s-1})$) and even \Threeopt\ with all the improvements described above still takes a lot of time to proceed.  However, \Threeopt\ is more robust when used in a combination with some other heuristic (see Section~\ref{sec:combined}).

\bigskip

It is worth noting that our extension of the pairwise (triple) interchange heuristic~\cite{Balas1991} is not typical.  Many papers~\cite{Aiex2005,Clemons2004,Murphey1998,Pasiliao2005,Robertson2001} consider another neighborhood:
$$
N_\text{\OptStarPlain{$k$}}(A) = \big\{ p_D(A, \rho) :\ D \subset \{ 1, 2, \ldots, s \}, |D| = 1 \text{ and $\rho$ moves at most $k$ elements} \big\} \ ,
$$
where $p_D$ is defined in~(\ref{eq:p_D}).  The size of such neighborhood is $|N_\text{\OptStarPlain{$k$}}(A)| = s \cdot \binom{n}{k} \cdot (k! - 1) + 1$ and the time complexity of one run of \OptStar{$k$} in the assumption $k \ll n$ is $O(s \cdot n^k \cdot k!)$, i.e., unlike \Kopt, it is not exponential with respect to the number of dimensions $s$ which is considered to be important by many researchers.  However, as it is stated in Section~\ref{sec:introduction}, we assume that $s$ is a small fixed constant and, thus, the time complexity of \Kopt\ is still reasonable.  At the same time, observe that $N_{k\text{-opt*}}(A) \subset N_\text{1-DV}(A)$ for any $k \le n$, i.e., \OneDV\ performs as good as \OptStar{$n$} with the time complexity of \OptStar{3}.  Only in the case of $k = 2$ the heuristic \OptStar{2} is faster in theory however it is known~\cite{Burkard1999} that the expected time complexity of AP is significantly less than $O(n^3)$ and, thus, the running times of \OptStar{2} and \OneDV\ are similar while \OneDV\ is definitely more powerful.  Because of this we do not consider \OptStar{2} in our comparison.

\subsection{Variable Depth Interchange (v-opt)}
\label{sec:vopt}

The \emph{Variable Depth Interchange} (\VDI) was first introduced by Balas and Saltzman for 3-AP as a heuristic based on the well known Lin-Kernighan heuristic for the traveling salesman problem~\cite{Balas1991}.  We provide here a natural extension \Vopt\ of the VDI heuristic for the $s$-dimensional case, $s \ge 3$, and then improve this extension.  Our computational experiments show that the improved version of \Vopt\ is superior to the natural extension of \VDI{} with respect to solution quality at the cost of a reasonable increase in running time.  In what follows, \Vopt{} refers to the improved version of the heuristic unless otherwise specified.

In~\cite{Balas1991}, the heuristic is described quite briefly.  Our contribution is not only in the extending, improving and analyzing it but also in a more detailed and, we believe, clearer explanation of it.  We describe the heuristic in a different way to the description provided in~\cite{Balas1991}, however, both versions of our algorithm are equal to \VDI{} in case of $s = 3$.  This fact was also checked by reproduction of the computational evaluation results reported in~\cite{Balas1991}.

%

Further we will use function $U(u, v)$ which returns a set of swaps between vectors $u$ and $v$.  The difference between the two versions of \Vopt{} is only in the $U(u, v)$ definition.  For the natural extension of \VDI{}, let $U(u, v)$ be a set of all the possible swaps (see~(\ref{eq:swap})) in at most one dimension between the vectors $u$ and $v$, where the coordinates in at most one dimension are swapped:
$$
U(u, v) = \big\{ swap(u, v, D) :\ D \subset \{ 1, 2, \ldots, s \} \text{ and } |D| \le 1 \big\} \ .
$$

For the improved version of \Vopt{}, let $U(u, v)$ be a set of all the possible swaps in at most $\lfloor s / 2 \rfloor$ dimensions between the vectors $v$ and $w$:
$$
U(u, v) = \big\{ swap(u, v, D) :\ D \subset \{ 1, 2, \ldots, s \} \text{ and } |D| \le s / 2 \big\} \ .
$$
The constraint $|D| \le s / 2$ guarantees that at least half of the coordinates of every swap are equal to the first vector coordinates.  The computational experiments show that removing this constraint increases the running time and decreases the average solution quality.

Let vector $\mu(u, v)$ be the minimum weight swap between vectors $u$ and $v$:
$$
\mu(u, v) = \argmin_{e \in U(u, v)} w(e) \ .
$$
Let $A$ be an initial assignment.

\begin{enumerate}
\item \label{item:vopt_mainloop} For every vector $c \in A$ do the rest of the algorithm.

\item \label{item:vopt_loopbegin} Initialize the \emph{total gain} $G = 0$, the \emph{best assignment} $A_\text{best} = A$, and a set of available vectors $L = A \setminus \{ c \}$.

\item \label{item:vopt_search} Find vector $m \in L$ such that $w(\mu(c, m))$ is minimized.  Set $v = \mu(c, m)$ and $\overline{v}_j = \{ c_j, m_j\} \setminus \{ v_j \}$ for every $1 \le j \le s$.
Now $v \in U(c, m)$ is the minimum weight swap of $c$ with some other vector $m$ in the assignment, and $\overline{v}$ is the complementary vector.

\item \label{item:vopt_stop_condition} Set $G = G + w(c) - w(v)$.  If now $G \le 0$, set $A = A_\text{best}$ and go to the next iteration (Step~\ref{item:vopt_mainloop}).

\item Mark $m$ as an unavailable for the further swaps: $L = L \setminus \{ m \}$.  Note that $c$ is already marked unavailable: $c \notin L$.

\item Replace $m$ and $c$ with $v$ and $\overline{v}$.  Set $c = \overline{v}$.

\item If $w(A) < w(A_\text{best})$, save the new assignment as the best one: $A_\text{best} = A$.

\item Repeat from Step~\ref{item:vopt_search} while the total gain is positive (see Step \ref{item:vopt_stop_condition}) and $L \neq \varnothing$.

\end{enumerate}

The heuristic repeats until no improvement is found during a run.  The time complexity of one run of \Vopt{} is $O(n^3 \cdot 2^{s-1})$.  The time complexity of the natural extension of \VDI{} is $O(n^3 \cdot s)$, and the computation experiments also show a significant difference between the running times of the improved and the natural extensions.  However, the solution quality of the natural extension for $s \ge 7$ is quite poor, while for the smaller values of $s$ it produces solutions similar to or even worse than \MDV{} solutions at the cost of much larger running times.

The neighborhood $N_\text{\VoptPlain}(A)$ is not fixed and depends on the MAP instance and initial assignment $A$.  The number of iterations (runs of Step~\ref{item:vopt_search}) of the algorithm can vary from $n$ to $n^2$.  Moreover, there is no guarantee that the algorithm selects a better assignment even if the corresponding swap is in $U(c, m)$.  Thus, we do not provide any results for the neighborhood of \Vopt{}.

\subsection{Combined Neighborhood}
\label{sec:combined}

We have already presented two types of neighborhoods in this paper, let us say \emph{dimensionwise} (Section~\ref{sec:dv}) and \emph{vectorwise} (Sections~\ref{sec:kopt} and~\ref{sec:vopt}).  The idea of the combined heuristic is to use the dimensionwise and the vectorwise neighborhoods togeather, combining them into so called Variable Neighborhood Search~\cite{Talbi2009}.  The combined heuristic improves the assignment by moving it into the local optimum with respect to the dimensionwise neighborhood, then it improves it by moving it to the local minimum with respect to the vectorwise neighborhood.  The procedure is repeated until the assignment occurs in the local minimum with respect to both the dimensionwise and the vectorwise neighborhoods.

More formally, the combined heuristic \Comb{DV}{opt} consists of a dimensionwise heuristic $DV$ (either \OneDV, \TwoDV\ or \MDV) and a vectorwise heuristic $opt$ (either \Twoopt, \Threeopt\ or \Vopt).  \Comb{DV}{opt} proceeds as follows.
\begin{enumerate}
	\item \label{item:combined_first} Apply the dimensionwise heuristic $A = DV(A)$.
	\item Repeat:
	\begin{enumerate}
		\item Save the assignment weight $x = w(A)$ and apply the vectorwise heuristic $A = opt(A)$.
		\item If $w(A) = x$ stop the algorithm.
		\item Save the assignment weight $x = w(A)$ and apply the dimensionwise heuristic $A = DV(A)$.
		\item If $w(A) = x$ stop the algorithm.
	\end{enumerate}
\end{enumerate}




Step~\ref{item:combined_first} of the combined heuristic is the hardest one.  Indeed, it is typical that it takes a lot of iterations to move a bad solution to a local minimum while for a good solution it takes just a few iterations.  Hence, the first of the two heuristics should be the most efficient one, i.e., it should perform quickly and produce a good solution.  In this case the dimensionwise heuristics are more efficient because, having approximately the same as vectorwise heuristics time complexity, they search much larger neighborhoods.  The fact that the dimensionwise heuristics are more efficient than the vectorwise ones is also confirmed by experimental evaluation (see Section~\ref{sec:experiments}).

It is clear that the neighborhood of a combined heuristic is defined as follows:
\begin{equation}
\label{eq:combined_neighborhood}
N_\text{\CombPlain{DV}{opt}}(A) = N_\text{DV}(A) \cup N_\text{opt}(A) \ ,
\end{equation}
where $N_\text{DV}(A)$ and $N_\text{opt}(A)$ are neighborhoods of the corresponding dimensionwise and vectorwise heuristics respectively.  To calculate the size of the neighborhood $N_\text{\CombPlain{DV}{opt}}(A)$ we need to find the size of the intersection of these neighborhoods.  Observe that
\begin{equation}
\label{eq:combined_intersection}
N_\text{DV}(A) \cap N_{k\text{-opt}}(A) = \big\{ p_D(A, \rho) :\ D \in \mathcal{D} \text{ and $\rho$ moves at most $k$ elements} \big\} \ ,
\end{equation}
where $p_D(A, \rho)$ is defined by~(\ref{eq:p_D}).  This means that, if $r_k$ is the number of permutations on $n$ elements which move at most $k$ elements, the intersection (\ref{eq:combined_intersection}) has size
\begin{equation}
\label{eq:combined_intersection_size}
|N_\text{DV}(A) \cap N_{k\text{-opt}}(A)| = |\mathcal{D}| \cdot (r_k - 1) + 1 \ .
\end{equation}
The number $r_k$ can be calculated as 
\begin{equation}
\label{eq:combined_r_k}
r_k = \sum_{i = 0}^k \binom{n}{i} \cdot d_i \ ,
\end{equation}
where $d_i$ is the number of derangements on $i$ elements, i.e., permutations on $i$ elements such that none of the elements appear on their places; $d_i = i! \cdot \sum_{m = 0}^i (-1)^m / m!$~\cite{Harris2008}.  For $k = 2$, $r_2 = 1 + \binom{n}{2}$; for $k = 3$, $r_3 = 1 + \binom{n}{2} + 2 \binom{n}{3}$.  From (\ref{eq:dv_neighborhood_size}), (\ref{eq:kopt_neighborhood_size}), (\ref{eq:combined_neighborhood}) and (\ref{eq:combined_intersection_size}) we immediately have
\begin{equation}
\left|N_\text{\CombPlain{DV}{\KoptPlain}}(A)\right| = 1 + |\mathcal{D}| \cdot (n! - 1) + \left[\sum_{i = 2}^k \binom{n}{i} N^i\right] - |\mathcal{D}| \cdot (r_k - 1) \ ,
\end{equation}
where $N^i$ and $r_k$ are calculated according to (\ref{eq:kopt_n_k}) and (\ref{eq:combined_r_k}) respectively.  Substituting the value of $k$, we have:
\begin{equation}
\label{eq:combined_neighborhood_size_2}
\left|N_\text{\CombPlain{DV}{\TwooptPlain}}(A)\right| = 1 + |\mathcal{D}| \cdot (n! - 1) + \binom{n}{2} (2^{s-1} - 1) - |\mathcal{D}| \cdot \binom{n}{2} \qquad \text{and}
\end{equation}
\begin{multline}
\label{eq:combined_neighborhood_size_3}
\left|N_\text{\CombPlain{DV}{\ThreeoptPlain}}(A)\right| = 1 + |\mathcal{D}| \cdot (n! - 1) + \binom{n}{2} (2^{s-1} - 1) \\
+ \binom{n}{3} (6^{s-1} - 3 \cdot 2^{s-1} + 2) - |\mathcal{D}| \cdot \left[\binom{n}{2} + 2 \binom{n}{3}\right]
\end{multline}

One can easily substitute $|\mathcal{D}| = s$, $|\mathcal{D}| = \binom{s}{2}$ and $|\mathcal{D}| = 2^{s-1} - 1$ to (\ref{eq:combined_neighborhood_size_2}) or (\ref{eq:combined_neighborhood_size_3}) to get the neighborhood sizes of \OneDVtwo, \TwoDVtwo, \MDVtwo, \OneDVthree, \TwoDVthree\ and \MDVthree.  We will only show the results for \MDVtwo:
%
\begin{multline}
\left|N_\text{\MDVtwoPlain}(A)\right| = 1 + (2^{s-1} - 1) \cdot (n! - 1) + \binom{n}{2} (2^{s-1} - 1) - (2^{s-1} - 1) \cdot \binom{n}{2} \\
= 1 + (2^{s-1} - 1) \cdot (n! - 1) \ ,
\end{multline}
i.e., $\left|N_\text{\MDVtwoPlain}(A)\right| = \left|N_\text{\MDVPlain}(A)\right|$\@.  Since $N_\text{\MDVPlain}(A) \subseteq N_\text{\MDVtwoPlain}(A)$ (see~(\ref{eq:combined_neighborhood})), we can conclude that $N_\text{\MDVtwoPlain}(A) = N_\text{\MDVPlain}(A)$.  Indeed, the neighborhood of \Twoopt\ can be defined as follows: 
$$
N_\text{\TwooptPlain} = \big\{ p_D(A, \rho) :\ D \subset \{ 2, 3, \ldots, s \} \text{ and } \rho \text{ swaps at most two elements} \big\} \ ,
$$
which is obviously a subset of $N_\text{\MDVPlain}(A)$ (see (\ref{eq:dv_neighborhood})).  Hence, the combined heuristic \MDVtwo{} is of no interest.

For other combinations the intersection (\ref{eq:combined_intersection}) is significantly smaller than both neighborhoods $N_\text{DV}(A)$ and $N_{\KoptPlain}(A)$ (recall that the neighborhood $N_\text{\VoptPlain}$ has a variable structure).  Indeed, $|N_\text{DV}(A)| \gg |N_\text{DV}(A) \cap N_{k\text{-opt}}(A)|$ because $|\mathcal{D}| \cdot (n! - 1) \gg |\mathcal{D}| \cdot (r_k - 1)$ for $k \ll n$.  Similarly, $|N_\text{\TwooptPlain}(A)| \gg |N_\text{DV}(A) \cap N_{k\text{-opt}}(A)|$ because $\binom{n}{2} (2^{s-1} - 1) \gg |\mathcal{D}| \cdot \binom{n}{2}$ if $|\mathcal{D}| \ll 2^{s-1}$, which is the case for \OneDV\ and \TwoDV\ if $s$ is large enough.  Finally, $|N_\text{\ThreeoptPlain}(A)| \gg |N_\text{DV}(A) \cap N_{k\text{-opt}}(A)|$ because $\binom{n}{2} (2^{s-1} - 1) + \binom{n}{3} (6^{s-1} - 3 \cdot 2^{s-1} + 2) \gg |\mathcal{D}| \cdot \left[ \binom{n}{2} + 2 \binom{n}{3} \right]$, which is true even for $|\mathcal{D}| = 2^{s-1}$, i.e., for \MDV.

The time complexity of the combined heuristic is $O(n^k \cdot k!^{s-1} + |\mathcal{D}| \cdot n^3)$ in case of $opt = \text{\KoptPlain}$ and $O(n^3 \cdot (2^{s-1} + |\mathcal{D}|))$ if $opt = \text{\VoptPlain}$.  The particular formulas are provided in the following table:

\begin{center}
\begin{tabular}{llll}
\toprule
 & \TwooptPlain & \ThreeoptPlain & \VoptPlain \\
\cmidrule(){1-4}
\OneDVPlain & $O(2^{s-1} \cdot n^2 + s \cdot n^3)$ & $O(6^{s-1} \cdot n^3)$ & $O(2^s \cdot n^3)$ \\
\TwoDVPlain & $O(2^{s-1} \cdot n^2 + s^2 \cdot n^3)$ & $O(6^{s-1} \cdot n^3)$ & $O(2^s \cdot n^3)$ \\
\MDVPlain & (no interest) & $O(6^{s-1} \cdot n^3)$ & $O(2^s \cdot n^3)$ \\
\bottomrule
\end{tabular}
\end{center}

Note that all the combinations with \Threeopt\ and \Vopt\ have equal time complexities; this is because the time complexities of \Threeopt\ and \Vopt\ are dominant.  Our experiments show that the actual running times of \Threeopt\ and \Vopt\ are really much higher then even the \MDV\ running time.  This means that the combinations of these heuristics with \MDV{} are approximately as fast as the combinations of these heuristics with light dimensionwise heuristics \OneDV\ and \TwoDV\@.  Moreover, as it was noticed above in this section, the dimensionwise heuristic, being executed first, simplifies the job for the vectorwise heuristic and, hence, the increase of the dimensionwise heuristic power may decrease the running time of the whole combined heuristic.  At the same time, the neighborhoods of the combinations with \MDV\ are significantly larger than the neighborhoods of the combinations with \OneDV\ and \TwoDV\@.  We can conclude that the `light' heuristics \OneDVthree, \TwoDVthree, \OneDVV\ and \TwoDVV\ are of no interest because the `heavy' heuristics \MDVthree\ and \MDVV, having the same theoretical time complexity, are more powerful and, moreover, outperformed the `light' heuristics in our experiments with respect to both solution quality and running time on average and in most of single experiments.




%
%
%


\subsection{Other algorithms}

Here we provide a list of some other MAP algorithms presented in the literature.

\begin{itemize}
\item A host of local search procedures and construction heuristics which often have some approximation guarantee (\cite{Bandelt2004,Burkard1996,Crama1992,Isler2005,Kuroki2007,Murphey1998} and some others) are proposed for special cases of MAP (usually with decomposable weights, see Section~\ref{sec:decomposable}) and exploit the specifics of these instances.  However, as it was stated in Section~\ref{sec:introduction}, we consider only the general case of MAP, i.e., all the algorithms included in this paper do not rely on any special structure of the weight matrix.

\item A number of construction heuristics are intended to generate solutions for general case MAP~\cite{Balas1991,Gutin2008,GK_MAP_Construction,Oliveira2004}.  While some of them are fast and low quality, like \Greedy, some, like \MaxRegret, are significantly slower but produce much better solutions.  A special class of construction heuristics, Greedy Randomized Adaptive Search Procedure (GRASP), was also investigated by many researchers~\cite{Aiex2005,Murphey1998,Oliveira2004,Robertson2001}.

\item Several metaheuristics, including a simulated annealing procedure~\cite{Clemons2004} and a memetic algorithm~\cite{Huang2006}, were proposed in the literature.  Metaheuristics are sophisticated algorithms intended to search for the near optimal solutions in a reasonably large time.  Proceeding for much longer than local search and being hard for theoretical analysis of the running time or the neighborhood, metaheuristics cannot be compared straightforwardly to local search procedures.

\item Some weak variations of \Twoopt\ are considered in~\cite{Aiex2005,Murphey1998,Pasiliao2005,Robertson2001}.  While our heuristic \Twoopt\ tries all possible recombinations of a pair of assignment vectors, i.e., $2^{s-1}$ combinations, these variations only try the swaps in one dimension at a time, i.e., $s$ combinations for every pair of vectors.  We have already decided that these variations have no practical interest, for details see Section~\ref{sec:kopt}.

\end{itemize}

\section{Test Bed}
\label{sec:testbed}

While the theoretical analysis can help in heuristic design, selection of the best approaches requires empirical evaluation~\cite{GK_Some_Theory,Rardin2001}.  In this section we discuss the test bed and in Section~\ref{sec:experiments} the experimental results are reported and discussed.

The question of selecting proper test bed is one of the most important questions in heuristic experimental evaluation~\cite{Rardin2001}.  While many researchers focused on instances with random independent weights (\cite{Andrijich2001,Balas1991,Krokhmal2007,Pasiliao2005} and some others) and random instances with predefined solutions~\cite{Clemons2004,Grundel2005,GK_MAP_Construction}, several more sophisticated models are of greater practical interest~\cite{Bandelt2004,Burkard1996,Crama1992,Frieze1981,Kuroki2007}.  There is also a number of papers which consider real-world and pseudo real-world instances~\cite{Bekker2005,Murphey1998,Pardalos2000b} but the authors of this paper suppose that these instances do not well represent all the instance classes and building a proper benchmark with the real-world instances is a subject for another research.

In this paper we group all the instance families into two classes: instances with independent weights (Section~\ref{sec:independent}) and instances with decomposable weights (Section~\ref{sec:decomposable}).  Later we show that the heuristics perform differently on the instances of these classes and, thus, this devision helps us in correct experimental analysis of the local search algorithms.

\subsection{Instances With Independent Weights}
\label{sec:independent}

One of the most studied class of instances for MAP is \emph{Random Instance Family}.  In \random{}, the weight assigned to a vector is a random uniformly distributed integral value in the interval $[a, b - 1]$.  \random{} instances were used in~\cite{Aiex2005,Andrijich2001,Balas1991,Pierskalla1968} and some others.

Since the instances are random and quite large, it is possible to estimate the average solution value for the Random Instance Family.  The previous research in this area~\cite{Krokhmal2007} show that if $n$ tends to infinity than the problem solution approaches the bound $an$, i.e., the minimal possible assignment weight (observe that the minimal assignment includes $n$ vectors of weight $a$).  Moreover, an estimation of the mean optimal solution is provided in~\cite{Grundel2004} but this estimation is not accurate enough for our experiments.  In~\cite{GK_Some_Theory} we prove that it is very likely that every big enough \random{} instance has at least one $an$-assignment, where \emph{$x$-assignment} means an assignment of weight $x$.

Let $\alpha$ be the number of assignments of weight $an$ and let $c = b - a$.  We would like to have an upper bound on the probability $\Pr(\alpha=0)$. Such an upper bound is given in the following theorem whose proof (see~\cite{GK_Some_Theory}) is based on the Extended Jansen Inequality (see Theorem 8.1.2 of~\cite{Alon2000}).

\begin{theorem}
\label{th:pr}
For any $n$ such that $n\ge 3$ and
\begin{equation}
\label{eq:cbound}
\left( \frac{n-1}{e} \right)^{s-1} \ge c \cdot 2^{\frac{1}{n-1}},
\end{equation}
we have $\Pr(\alpha = 0) \le e^{-\frac{1}{2 \sigma}}$, where
$\sigma = \sum\limits_{k = 1}^{n - 2} \frac{\binom{n}{k} \cdot c^k}{\left[n \cdot (n - 1) \cdots (n - k + 1) \right]^{s - 1}}.$
\end{theorem}

The lower bounds of $\Pr(\alpha > 0)$ for different values of $s$ and $n$ and for $b - a = 100$, are reported below.

\begin{center}
\noindent\begin{tabular*}{0.95\textwidth}{@{\extracolsep{\fill}} *5{c}}
\toprule{}
$s = 4$
    & $s = 5$
    & $s = 6$
    & $s = 7$ \\

\cmidrule(){1-4}

\begin{tabular}[t]{r @{\quad} l}
$n$ & $\Pr(\alpha > 0)$ \\
\cmidrule(){1-2}
15 & 0.575 \\
20 & 0.823 \\
25 & 0.943 \\
30 & 0.986 \\
35 & 0.997 \\
40 & 1.000 \\
\end{tabular}
&
\begin{tabular}[t]{r @{\quad} l}
$n$ & $\Pr(\alpha > 0)$ \\
\cmidrule(){1-2}
10 & 0.991 \\
11 & 0.998 \\
12 & 1.000 \\
\end{tabular}
&
\begin{tabular}[t]{r @{\quad} l}
$n$ & $\Pr(\alpha > 0)$ \\
\cmidrule(){1-2}
8 & 1.000 \\
\end{tabular}
&
\begin{tabular}[t]{r @{\quad} l}
$n$ & $\Pr(\alpha > 0)$ \\
\cmidrule(){1-2}
7 & 1.000 \\
\end{tabular} \\
\bottomrule{}
\end{tabular*}
\end{center}

One can see that a 4-AP \random{} instance has an $(an)$-assignment with the probability which is very close to 1 if $n \ge 40$; a 5-AP instance has an $(an)$-assignment with probability very close to 1 for $n \ge 12$, etc.; so, the optimal solutions for all the \random{} instances used in our experiments (see Section~\ref{sec:experiments}) are very likely to be of weight $an$.  For $s = 3$ Theorem~\ref{th:pr} does not provide a good upper bound, but we are able to use the results from Table~II in~\cite{Balas1991} instead.  Balas and Saltzman report that in their experiments the average optimal solution of 3-AP for \random{} instances reduces very quickly and has a small value even for $n = 26$.  Since the smallest \random{} instance we use in our experiments has size $n = 150$, we assume that all 3-AP \random{} instances considered in our experiment are very likely to have an $an$-assignment.

Useful results can also be obtained from (11) in~\cite{Grundel2004} which is an upper bound for the average optimal solution.  Grundel, Oliveira and Pardalos~\cite{Grundel2004} consider the same instance family except the weights of the vectors are real numbers uniformly distributed in the interval $[a, b]$. However the results from~\cite{Grundel2004} can be extended to our discrete case.  Let $w'(e)$ be a real weight of the vector $e$ in a continuous instance.  Consider a discrete instance with $w(e) = \lfloor w'(e) \rfloor$ (if $w'(e) = b$, set $w(e) = b - 1$).  Note that the weight $w(e)$ is a uniformly distributed integer in the interval $[a, b - 1]$.  The optimal assignment weight of this instance is not larger than the optimal assignment weight of the continuous instance and, thus, the upper bound for the average optimal solution for the discrete case is correct.

In fact, the upper bound $\bar{z}^*_u$ (see~\cite{Grundel2004}) for the average optimal solution is not accurate enough.  For example, $\bar{z}^*_u \approx an + 6.9$ for $s = 3$, $n = 100$ and $b - a = 100$, and $\bar{z}^*_u \approx an + 3.6$ for $s = 3$, $n = 200$ and $b - a = 100$, so it cannot be used for $s = 3$ in our experiments. The upper bound $\bar{z}^*_u$ gives a better approximation for larger values of $s$, e.g., $\bar{z}^*_u \approx an + 1.0$ for $s = 4$, $n = 40$ and $b - a = 100$, however, Theorem~\ref{th:pr} provides stronger results ($\Pr(\alpha > 0) \approx 1.000$ for this case).

\bigskip

Another class of instances with almost independent weights is \emph{GP Instance Family} which contains pseudo-random instances with predefined optimal solutions.  \GP{} instances are generated by an algorithm produced by Grundel and Pardalos~\cite{Grundel2005}.  The generator is naturally designed for $s$-AP for arbitrary large values of $s$ and $n$.  However, it is relatively slow and, thus, it was impossible to generate large \GP{} instances.  Nevertheless, this is what we need since finally we have both small (\GP) and large (\random) instances with independent weights with known optimal solutions.

\subsection{Instances With Decomposable Weights}
\label{sec:decomposable}

In many cases it is not easy to define a weight for an $s$-tuple of objects but it is possible to define a relation between every pair of objects from different sets.  In this case one should use \emph{decomposable weights}~\cite{Spieksma2000} (or \emph{decomposable costs}), i.e., the weight of a vector $e$ should be defined as follows:
\begin{equation}
\label{eq:decomposable_w}
w(e) = f\left(d^{1, 2}_{e_1, e_2}, d^{1, 3}_{e_1, e_3}, \ldots, d^{s-1,s}_{e_{s-1}, e_s}\right) \ ,
\end{equation}
where $d^{i,j}$ is a distance matrix between the sets $X_i$ and $X_j$ and $f$ is some function.

The most natural instance family with decomposable weights is \clique, which defines the function $f$ as the sum of all arguments:
\begin{equation}
\label{eq:clique_w}
w_\text{c}(e) = \sum_{i=1}^{n-1} \sum_{j=i+1}^{n} d^{i,j}_{e_i, e_j} \ .
\end{equation}
The \clique\ instance family was investigated in~\cite{Bandelt2004,Crama1992,Frieze1981} and some others.  It was proven~\cite{Crama1992} that MAP restricted to \clique\ instances remains NP-hard.

A special case of \clique\ is \emph{Geometric Instance Family}.  In \geometric, the sets $X_1$, $X_2$, \ldots, $X_s$ correspond to sets of points in Euclidean space, and the distance between two points $u \in X_i$ and $v \in X_j$ is defined as Euclidean distance; we consider the two dimensional Euclidean space:
$$
d_\text{g}(u, v) = \sqrt{(u_x - v_x)^2 + (u_y - v_y)^2} \ .
$$
It is proven~\cite{Spieksma1996} that the \geometric\ instances are NP-hard to solve for $s = 3$ and, thus, \geometric\ is NP-hard for every $s \ge 3$.

In this paper, we propose a new special case of the decomposable weights, \squareroot.  It is a modification of the \clique\ instance family.  Assume we have $s$ radars and $n$ planes and each radar observes all the planes.  The problem is to assign signals which come from different radars to each other.  It is quite natural to define a distance function between each pair of signals from different radars, and for a set of signals which correspond to one plane the sum of the distances should be small so (\ref{eq:clique_w}) is a good choice.  However, it is not actually correct to minimize the total distance between the assigning signals but one should also ensure that none of these distances is too large.  Same requirements appear in a number of other applications.  We propose a weight function which can leads to both small total distance between the assigned signals and small dispersion of the distances:
\begin{equation}
w_\text{sq}(e) = \sqrt{\sum_{i=1}^{n-1} \sum_{j=i+1}^{n} \left(d^{i,j}_{e_i, e_j}\right)^2 } \ .
\end{equation}
Similar approach is used in~\cite{Kuroki2007} though they do not use square root, i.e., a vector weight is just a sum of squares of the edge weights in a clique.  In addition, the edge weights in~\cite{Kuroki2007} are calculated as distances between some nodes in a Euclidean space.

Another special case of the decomposable weights, \product, is studied in~\cite{Burkard1996}.  Burkard et al. consider 3-AP and define the weight $w(e)$ as $w(e) = a^1_{e_1} \cdot a^2_{e_2} \cdot a^3_{e_3}$, where $a^1$, $a^2$ and $a^3$ are random vectors of positive numbers.
It is easy to show that the \product\ weight function can be represented in the form~(\ref{eq:decomposable_w}).
It is proven that the minimization problem for the \product\ instances is NP-hard in case $s = 3$ and, thus, it is NP-hard for every $s \ge 3$.

\section{Computational Experimentation}
\label{sec:experiments}

In this section, the results of empirical evaluation are reported and discussed.  The experiments were conducted for the following instances (for instance family definitions see Section~\ref{sec:testbed}):
\begin{itemize}
    \item \random{} instances where each weight was randomly chosen in $\{ 1, 2, \ldots, 100 \}$, i.e., $a = 1$ and $b = 101$.  According to Subsection~\ref{sec:independent}, the optimal solutions of all the considered \random\ instances are very likely to be $an = n$.

    \item \GP{} instances with predefined optimal solutions (see Section~\ref{sec:independent}).

    \item \clique\ and \squareroot\ instances, where the weight of each edge in the graph was randomly selected from $\{ 1, 2, \ldots, 100 \}$.  Instead of the optimal solution value we use the best known solution value.

    \item \geometric\ instances, where both coordinates of every point were randomly selected from $\{ 1, 2, \ldots, 100 \}$.  The distances between the points are calculated precisely while the weight of a vector is rounded to the nearest integer.  Instead of the optimal solution value we use the best known solution value.

    \item \product\ instances, where every value $a_i^j$ was randomly selected from $\{ 1, 2, \ldots, 10 \}$.  Instead of the optimal solution value we use the best known solution value.

\end{itemize}

An instance name consists of three parts: the number $s$ of dimensions, the type of the instance (`gp' for \GP{}, `r' for \random{}, `c' for \clique{}, `g' for \geometric, `p' for \product\ and 'sr' for \squareroot), and the size $n$ of the instance.  For example, \texttt{5r40} means a five dimensional \random\ instance of size 40.  For every combination of instance size and type we generated 10 instances, using the number $seed = s + n + i$ as a seed of the random number sequences, where $i$ is an index of the instance of this type and size, $i \in \{ 1, 2, \ldots, 10 \}$.  Thereby, every experiment is conducted for 10 different instances of some fixed type and size, i.e., every number reported in the tables below is average for 10 runs for 10 different instances.

The sizes of all but \GP\ instances are selected such that every algorithm could process them all in approximately the same time.  The \GP\ instances are included in order to examine the behavior of the heuristics on smaller instances (recall that \GP\ is the only instance set for which we know the exact solutions for small instances).


All the heuristics are implemented in Visual C++.  The evaluation platform is based on AMD Athlon 64 X2 3.0~GHz processor.

Further, the results of the experiments of three different types are provided and discussed:
\begin{itemize}
\item In Subsection~\ref{sec:pure}, the local search heuristics are applied to the assignments generated by some construction heuristic.  These experiments allow us to exclude several local searches from the rest of the experiments, however, the comparison of the results is complicated because of the significant difference in both the solution quality and the running time.

\item In Subsection~\ref{sec:metaheuristics}, two simple metaheuristics are used to equate the running times of different heuristics.  This is done by varying of number of iterations of the metaheuristics.

\item In Subsection~\ref{sec:heuristics_comparison}, the results of all the discussed approaches are gathered in two tables to find the most successful solvers for the instance with independent and decomposable weights for every particular running time.
\end{itemize}

\subsection{Pure Local Search Experiments}
\label{sec:pure}

First, we run every local search heuristic for every instance exactly once.  The local search is applied to solutions generated with one of the following construction heuristics:
\begin{enumerate}
\item \Trivial, which was first mentioned in~\cite{Balas1991} as \emph{Diagonal}.  \Trivial\ construction heuristic simply assigns $A_j^i = i$ for every $i = 1, 2, \ldots, n$ and $j = 1, 2, \ldots, s$.

\item \Greedy\ heuristic was discussed in many papers, see, e.g.~\cite{Balas1991,Burkard1996,Gutin2008,GK_Greedy_Chapter,GK_Some_Theory,GK_MAP_Construction}.  It was proven~\cite{Gutin2008} that in the worst case \Greedy\ produces the unique worst solution; however, it was shown~\cite{GK_Greedy_Chapter} that in some cases \Greedy\ may be a good selection as a fast and simple heuristic.

\item \MaxRegret\ was discussed in a number of papers, see, e.g.,~\cite{Balas1991,Burkard1996,Gutin2008,GK_MAP_Construction,Robertson2001}.  As for \Greedy, it is proven~\cite{Gutin2008} that in the worst case \MaxRegret\ produces the unique worst solution however many researchers~\cite{Balas1991,GK_MAP_Construction} noted that \MaxRegret\ is quite powerful in practice.

\item \ROM\ was first introduced in~\cite{Gutin2008} as a heuristic of a large domination number.  On every iteration, the heuristic calculates the total weight for every set of vectors with the fixed first two coordinates: $M_{i,j} = \sum_{e \in X, e_1 = i, e_2 = j} w(e)$.  Then it solves a 2-AP for the weight matrix $M$ and reorders the second dimension of the assignment according to this solution and the first dimension of the assignment.  The procedure is repeated recursively for the subproblem where the first dimension is excluded.  For details see~\cite{Gutin2008,GK_MAP_Construction}.

\end{enumerate}

We will begin our discussion from the experiments started from trivial assignments.  The results reported in Tables~\ref{tab:ls_solutions_from_trivial} and~\ref{tab:ls_times_from_trivial} are averages for 10 experiments since every row of these tables corresponds to 10 instances of some fixed type and size but of different seed values (see above).  The tables are split into two parts; the first part contains only the instances with independent weights (\GP\ and \random) while the second part contains only the instances with decomposable weights (\clique, \geometric, \product\ and \squareroot).  The average values for different instance families and numbers of dimensions are provided at the bottom of each part of each table.  The tables are also split vertically according to the classes of heuristics.  The winner in every row and every class of heuristics is underlined.

The value of the solution error is calculated as $\big(w(A) / w(A_\text{best}) - 1\big) \cdot 100\%$, where $A$ is the obtained assignment and $A_\text{best}$ is the optimal assignment (or the best known one, see above).

\bigskip

In the group of the vectorwise heuristics the most powerful one is definitely \Threeopt.  \Vopt{} outperforms it only in a few experiments, mostly three dimensional ones (recall that the neighborhood of \Kopt\ increases exponentially with the increase of the number of dimensions $s$).  As it was expected, \Twoopt\ never outperforms \Threeopt\ since $N_\text{2-opt} \subset N_\text{3-opt}$ (see Section~\ref{sec:kopt}).  The tendencies for the independent weight instances and for the decomposable weight instances are similar; the only difference which is worth to note is that all but \Vopt\ heuristics of this group solve the \product\ instances very well.  Note that the dispersion of the weights in \product\ instances is really high and, thus, \Vopt, which minimizes the weight of only one vector in every pair of vectors while the weight of the complementary vector may increase arbitrary, cannot be efficient for them.

As one can expect, \MDV\ is more successful than \TwoDV\ and \TwoDV\ is more successful than \OneDV\ with respect to the solution quality (obviously, all the heuristics of this group perform equally for 3-AP and \TwoDV\ and \MDV\ are also equal for 4-AP, see Section~\ref{sec:dv}).  However, for the instances with decomposable weights all the dimensionwise heuristics perform very similarly and even for the large $s$, \MDV\ is not significantly more powerful than \OneDV\ or \TwoDV\ which means that in case of decomposable instances the most efficient iterations are when $|D| = 1$.  We can assume that if $c$ is the number of edges connecting the fixed and unfixed parts of the clique, then an iteration of a dimensionwise heuristic is rather efficient when $c$ is small.  Observe that, e.g., for \clique\ the diversity of values in the weight matrix $[M_{i, j}]_{n \times n}$ (see~(\ref{eq:M})) decreases with the increase of the number $c$ and, hence, the space for optimization on every iteration is decreasing.  Observe also that in the case $c = 1$ the iteration leads to the optimal match between the fixed and unfixed parts of the assignment vectors.

All the combined heuristics show improvements in the solution quality over each of their components, i.e., over both corresponding vectorswise and dimensionwise local searches.  In particular, \OneDVtwo\ outperforms both \Twoopt\ and \OneDV, \TwoDVtwo\ outperforms both \Twoopt\ and \TwoDV, \MDVthree\ outperforms both \Threeopt\ and \MDV\ and \MDVV\ outperforms both \Vopt\ and \MDV\@.  Moreover, \MDVthree\ is significantly faster than \Threeopt\ and \MDVV\ is significantly faster than \Vopt.  Hence, we will not discuss the single heuristics \Threeopt\ and \Vopt\ in the rest of the paper.  The heuristics \OneDVtwo\ and \TwoDVtwo, obviously, perform equally for 3-AP instances.

While for the instances with independent weights the combination of the dimensionwise heuristics with the vectorwise ones significantly improves the solution quality, it is not the case for the instances with decomposable weights (observe that \OneDV\ performs almost as well as the most powerful heuristic \MDVthree) which shows the importance of the instances division.  We conclude that the vectorwise neighborhoods are not efficient for the instances with decomposable weights.

\bigskip

Next we conducted the experiments starting from the other construction heuristics.  But first we compared the construction heuristics themselves, see Table~\ref{tab:construction}.  It is not surprising that \Trivial\ produces the worst solutions.  However, one can see that \Trivial\ outperforms \Greedy\ and \MaxRegret\ for every \product\ instance.  The reason is in the extremely high dispersion of the weights in \product.  Both \Greedy\ and \MaxRegret\ construct the assignments by adding new vectors to it.  The decision which vector should be added does not depend (or does not depend enough in case of \MaxRegret) on the rest of the vectors and, thus, at the end of the procedure only the vectors with huge weights are available.  For other instance families, \Greedy, \MaxRegret\ and \ROM\ perform similarly though the running time of the heuristics is very different.  \MaxRegret\ is definitely the slowest construction heuristic; \Greedy\ is very fast for the \random\ instances (this is because of the large number of vectors of the weight $a$ and the implementation features, see~\cite{GK_MAP_Construction} for details) and relatively slow for the rest of the instances; \ROM's running time almost does not depend on the instance and is constantly moderate.

Starting from \Greedy\ (Table~\ref{tab:greedy}) significantly improves the solution quality.  This mostly influenced the weakest heuristics, e.g., \Twoopt\ average error decreased in our experiments from 59\% and 20\% to 15\% and 6\% for independent and decomposable weights respectively, though, e.g., the most powerful heuristic \MDVthree\ error also noticeably decreased (from 2.8\% and 5.8\% to 2.0\% and 2.5\%).  As regards the running time, \Greedy\ is slower than most of the local search heuristics and, thus, the running times of all but \MDVthree\ and \MDVV\ heuristics are very similar.  The best of the rest of the heuristics in this experiment is \MDV\ though \OneDVtwo\ and \TwoDVtwo\ perform similarly.

Starting from \MaxRegret\ improves the solution quality even more but at the cost of very large running times.  In this case the difference in the running time of the local search heuristics almost disappears and \MDVthree, the best one, reaches the average error values 1.3\% and 2.2\% for independent and decomposable weights respectively.  Starting from \ROM\ improves the quality only for the worst heuristics.  This is probably because all the best heuristics contain \MDV\ which does a good vectorwise optimization (recall that \ROM\ exploits a similar to the dimensionwise neighborhood idea).  At the same time, starting from \ROM\ increases the running time of the heuristics significantly; the results for both \MaxRegret\ and \ROM\ are excluded from the paper; one can find them on the web~\cite{Karapetyan}.

It is clear that the construction heuristics are quite slow comparing to the local search and we should answer the following question: is it worth to spend so much time on the initial solution construction or there is some way to apply local search several times in order to improve the assignments iteratively?  It is known that the algorithms which apply local search several times are called metaheuristics.  There is a number of different metaheuristic approaches such as tabu search or memetic algorithms, but this is not the subject of this paper.  In what follows, we are going to use two simple metaheuristics, \Chain\ and \Multichain.

\subsection{Experiments With Metaheuristics}
\label{sec:metaheuristics}

It is obvious that there is no sense in applying a local search procedure to one solution several times because the local search moves the solution to a local minimum with respect to its neighborhood, i.e., the second exploration of this neighborhood is useless.  In order to apply the local search several times, one should perturb the solution obtained on the previous iteration.  This idea immediately brings us to the first metaheuristic, let us say \Chain:
\begin{enumerate}
\item Initialize an assignment $A$;
\item Set $A_\text{best} = A$;
\item Repeat:
\begin{enumerate}
	\item Apply local search $A = LS(A)$;
	\item If $w(A) < w(A_\text{best})$ set $A_\text{best} = A$;
	\item Perturb the assignment $A = Perturb(A)$.
\end{enumerate}
\end{enumerate}

In this algorithm we use two subroutines, $LS(A)$ and $Perturb(A)$.  The first one is some local search procedure and the second one is an algorithm which removes the given assignment from the local minimum by a random perturbation of it.  The perturbation should be strong enough such that the assignment will not come back to the previous position on the next iteration every time though it should not be too strong such that the results of the previous search would be totally destroyed.  Our perturbation procedure selects $p = \lceil n / 25 \rceil + 1$ vectors in the assignment and perturbs them randomly.  In other words, $Perturb(A)$ is just a random move of the \Opt{p} heuristic.  The parameters of the procedure are obtained empirically.

One can doubt if \Chain\ is good enough for large running times and, thus, we introduce a little bit more sophisticated metaheuristic, \Multichain.  Unlike \Chain, \Multichain\ maintains several assignments on every iteration:
\begin{enumerate}
\item Initialize assignment $A_\text{best}$;
\item Set $P = \varnothing$ and repeat the following $c (c + 1) / 2$ times:\\
$P = P \cup \{ LS(Perturb(A_\text{best})) \}$\\
(recall that $Perturb(A)$ produces a different assignment every time);
\item Repeat:
\begin{enumerate}
	\item Save the best $c$ assignments from $P$ into $C_1, C_2, \ldots, C_c$ such that $w(C_i) \le w(C_{i+1})$;
	\item If $w(C_1) < w(C_\text{best})$ set $A_\text{best} = C_1$.
	\item Set $P = \varnothing$ and for every $i = 1, 2, \ldots, c$ repeat the following $c - i + 1$ times: $P = P \cup \{ LS(Perturb(C_i)) \}$.
\end{enumerate}
\end{enumerate}

The parameter $c$ is responsible for the power of \Multichain; we use $c = 5$ and, thus, the algorithm performs $c (c + 1) / 2 = 15$ local searches on every iteration.

\bigskip

The results of the experiments with \Chain\ running for 5 and 10 seconds are provided in Tables~\ref{tab:chain5} and~\ref{tab:chain10} respectively.  The experiments are repeated for three construction heuristics, \Trivial, \Greedy\ and \ROM\@.  It was not possible to include \MaxRegret\ in the comparison because it takes much more than 10 seconds for some of the instances.

The diversity in solution quality of the heuristics decreased with the usage of a metaheuristic.  This is because the fast heuristics are able to repeat more times than the slow ones.  Note that \MDVthree, which is the most powerful single heuristic, is now outperformed by other heuristics.  The most successful heuristics for the instances with independent and decomposable weights are \MDVV\ and \OneDV\ respectfully, though \OneDVtwo\ and \TwoDVtwo\ are slightly more successful than \MDVV\ for the \GP\ instances.  This result also holds for \Multichain, see Tables~\ref{tab:multichain5} and~\ref{tab:multichain10}.  The success of \OneDV\ confirms again that a dimensionwise heuristic is most successful when $|D| = 1$ if the weights are decomposable and that it is more efficient to repeat these iterations many times rather than try $|D| > 1$.  For the explanation of this phenomenon see Section~\ref{sec:pure}.  The success of \OneDVtwo\ and \TwoDVtwo\ for \GP\ means existence of a certain structure in the weight matrices of these instances.

One can see that the initialization of the assignment is not crucial for the final solution quality.  However, using \Greedy\ instead of \Trivial\ clearly improves the solutions for almost every instance and local search heuristic.  In contrast to \Greedy, using of \ROM\ usually does not improve the solution quality.  It only influences \Twoopt\ which is the only pure vectorwise local search in the comparison (recall that \ROM\ has a dimensionwise structure and, thus, it is good in combination with vectorwise heuristics).

The \Multichain\ metaheuristic, given the same time, obtains better results than \Chain.  However, \Multichain\ fails for some combinations of slow local search and hard instance because it is not able to complete even the first iteration in the given time.  \Chain, having much easier iterations, do not have this disadvantage.

Giving more time to a metaheuristic also improves the solution quality.  Therefore, one is able to obtain high quality solutions using metaheuristics with large running times.

\subsection{Solvers Comparison}
\label{sec:heuristics_comparison}

To compare all the heuristics and metaheuristics discussed in this paper we produced Tables~\ref{tab:heuristics_independent} and~\ref{tab:heuristics_decomposable}.  These tables indicate which heuristics should be chosen to solve particular instances in the given time limitations.  Several best heuristics are selected for every combination of the instance and the given time.  A heuristic is included in the table if it was able to solve the problem in the given time, and if its solution quality is not worse than $1.1 \cdot w(A_\text{best})$ and its running time is not larger than $1.1 \cdot t_\text{best}$, where $A_\text{best}$ is the best assignment produced by the considered heuristics and $t_\text{best}$ is the time spent to produce $A_\text{best}$.

The following information is provided for every solver in Tables~\ref{tab:heuristics_independent} and~\ref{tab:heuristics_decomposable}:
\begin{itemize}
	\item Metaheuristic type (\textbf{C} for \Chain{}, \textbf{MC} for \Multichain{} or empty if the experiment is single).
	\item Local search procedure (\Twoopt, \OneDV, \TwoDV, \MDV, \OneDVtwo, \TwoDVtwo, \MDVthree\, \MDVV\ or empty if no local search was applied to the initial solution).
	\item Construction heuristic the experiment was started with (\heuristic{Gr}, \heuristic{M-R} or empty if the assignment was initialized by \Trivial).
	\item The solution error in percent.
\end{itemize}

The following solvers were included in this experiment:
\begin{itemize}
	\item Construction heuristics \Greedy, \MaxRegret\ and \ROM.

	\item Single heuristics \Twoopt, \OneDV, \TwoDV, \MDV, \OneDVtwo, \TwoDVtwo, \MDVthree\ and \MDVV started from either \Trivial, \Greedy, \MaxRegret\ or \ROM.

	\item \Chain\ and \Multichain\ metaheuristics for either \Twoopt, \OneDV, \TwoDV, \MDV, \OneDVtwo, \TwoDVtwo, \MDVthree\ or \MDVV\ and started from either \Trivial, \Greedy, \MaxRegret\ or \ROM\@.  The metaheuristics proceeded until the given time limitations.
\end{itemize}

Note that for certain instances we exclude duplicating solvers (recall that all the dimensionwise heuristics perform equally for 3-AP as well as \TwoDV\ and \MDV\ perform equally for 4-AP, see Section~\ref{sec:dv}).  The common rule is that we leave \MDV\ rather than \TwoDV\ and \TwoDV\ rather than \OneDV\@.  For example, if the list of successful solvers for some 3-AP instance contains \textbf{C}~\OneDVPlain~Gr, \textbf{C}~\TwoDVPlain~Gr and \textbf{C}~\MDVPlain~Gr, then only \textbf{C}~\MDVPlain~Gr will be included in the table.  This is also applicable to the combined heuristics, e.g, having \OneDVtwoPlain~R and \TwoDVtwoPlain~R for a 3-AP instance, we include only \TwoDVtwoPlain~R in the final results.

The last row in every table indicates the heuristics which are the most successful on average, i.e., the heuristics which can solve all the instances with the best average results.

\bigskip

Single construction heuristics are not presented in the tables; single local search procedures appear only for the small allowed times when all other heuristics take more time to run; the most of the best solvers are the metaheuristics.  \Multichain\ seems to be more suitable than \Chain\ for large running times; however, \Multichain\ does not appear for the instances with small $n$.  This is probably because the power of the perturbation degree increases with the decrease of the instance size (note that $perturb(A)$ perturbs at least two vectors in spite of $n$).

The most successful heuristics for the assignment initialization are \Trivial\ and \Greedy; \Trivial\ is useful rather for small running times.  \MaxRegret\ and \ROM\ appear only a few times in the tables. 

The success of a local search depends on the instance type.  The most successful local search heuristic for the instance with independent weights is definitely \MDVV.  The \MDV\ heuristic also appears several times in Table~\ref{tab:heuristics_independent}, especially for the small running times.  For the instances with decomposable weights, the most successful are the dimensionwise heuristics and, in particular, \OneDV.

\section{Conclusions}

Several neighborhoods are generalized and discussed in this paper.  An efficient approach of joining different neighborhoods is successfully applied; the yielded heuristics showed that they combine the strengths of their components.  The experimental evaluation for a set of instances of different types show that there are several superior heuristic approaches suitable for different kinds of instances and running times.  Two kinds of instances are selected: instances with independent weights and instances with decomposable weights.  The first ones are better solvable by a combined heuristic \MDVV; the second ones are better solvable by \OneDV\@.  In both cases, it is good to initialize the assignment with the \Greedy\ construction heuristic if there is enough time; otherwise one should use a trivial assignment as the initial one.  The results can also be significantly improved by applying metaheuristic approaches for as log as possible.

Thereby, it is shown in the paper that metaheuristics applied to the fast heuristics dominate slow heuristics and, thus, further research of some more sophisticated metaheuristics such as memetic algorithms is of interest.



\bigskip

{\small
\bibliography{MapLocalSearchPaper}{}
\bibliographystyle{plain}
}

%


\begin{table}[ht] \centering
\scriptsize
\caption{Construction heuristics comparison.}
\label{tab:construction}
\begin{tabular}{@{} l @{\hspace{0.5em}} r @{\hspace{0.5em}} c @{\hspace{2em}} *{3}{r @{\hspace{0.5em}}} r @{} c @{\hspace{2em}} *{3}{r @{\hspace{0.5em}}} r @{}}
\toprule
 &  &  & \multicolumn{4}{c}{Solution error, \%} &  & \multicolumn{4}{c}{Running times, ms}\\
\cmidrule(){4-7}
\cmidrule(){9-12}
Inst. & Best &  & Trivial & Greedy & Max-Regret & ROM &  & Trivial & Greedy & Max-Regret & ROM\\
\cmidrule(){1-12}
3gp100 & 504.4& & 157 & \underline{6} & 6 & 10 & & \underline{0} & 40 & 799 & 9\\
3r150 & 150.0& & 4\,997 & 54 & \underline{29} & 34 & & \underline{0} & 14 & 4\,253 & 26\\
4gp30 & 145.2& & 158 & 9 & 9 & \underline{2} & & \underline{0} & 35 & 206 & 7\\
4r80 & 80.0& & 4\,985 & 74 & \underline{49} & 76 & & \underline{0} & 12 & 27\,285 & 278\\
5gp12 & 66.2& & 147 & 13 & \underline{9} & 9 & & \underline{0} & 6 & 36 & 2\\
5r40 & 40.0& & 4\,911 & 159 & \underline{116} & 169 & & \underline{0} & 6 & 37\,214 & 686\\
\bottomrule
\end{tabular}
\end{table}

\begin{table}[ht] \centering
\scriptsize
\caption{Local search heuristics started from Trivial.}
\label{tab:ls_solutions_from_trivial}
\begin{tabular}{@{} l@{\hspace{0.5em}}r@{\hspace{0.5em}} c @{\hspace{2em}} *{2}{r @{\hspace{0.5em}}} r @{} c @{\hspace{2em}} *{2}{r @{\hspace{0.5em}}} r @{} c @{\hspace{2em}} *{3}{r @{\hspace{0.5em}}} r @{}}
\toprule
 &  &  & \multicolumn{12}{c}{Solution error, \%}\\
\cmidrule(){4-15}
Inst. & Best &  & \TwooptPlain & \ThreeoptPlain & \VoptPlain &  & \OneDVPlain & \TwoDVPlain & \MDVPlain &  & \OneDVtwoPlain & \TwoDVtwoPlain & \MDVthreePlain & \MDVVPlain\\
\cmidrule(){1-15}
3gp100 & 504.4& & 19.6 & \underline{10.0} & 19.8 & & \underline{4.9} & \underline{4.9} & \underline{4.9} & & 4.9 & 4.9 & \underline{4.6} & 4.9\\
3r150 & 150.0& & 134.5 & 16.0 & \underline{1.5} & & \underline{2.4} & \underline{2.4} & \underline{2.4} & & 2.4 & 2.4 & 2.1 & \underline{0.7}\\
4gp30 & 145.2& & 17.4 & \underline{4.2} & 13.4 & & 11.1 & \underline{7.9} & \underline{7.9} & & 10.7 & 7.9 & \underline{4.2} & 7.5\\
4r80 & 80.0& & 115.0 & 7.3 & \underline{2.0} & & 20.5 & \underline{11.5} & \underline{11.5} & & 18.9 & 11.5 & 4.1 & \underline{1.6}\\
5gp12 & 66.2& & 10.6 & \underline{2.1} & 8.5 & & 12.5 & \underline{6.9} & \underline{6.9} & & 10.1 & 6.9 & \underline{1.8} & 6.9\\
5r40 & 40.0& & 104.5 & 4.3 & \underline{3.8} & & 63.0 & \underline{34.3} & \underline{34.3} & & 47.3 & 34.3 & \underline{3.5} & 5.3\\
\bottomrule
\end{tabular}
\end{table}

\begin{table}[ht] \centering
\scriptsize
\caption{Local search heuristics started from Trivial.}
\label{tab:ls_times_from_trivial}
\begin{tabular}{@{} l@{\hspace{0.5em}} c @{\hspace{2em}} *{2}{r @{\hspace{0.5em}}} r @{} c @{\hspace{2em}} *{2}{r @{\hspace{0.5em}}} r @{} c @{\hspace{2em}} *{3}{r @{\hspace{0.5em}}} r @{}}
\toprule
 &  & \multicolumn{12}{c}{Running time, ms}\\
\cmidrule(){3-14}
Inst. &  & \TwooptPlain & \ThreeoptPlain & \VoptPlain &  & \OneDVPlain & \TwoDVPlain & \MDVPlain &  & \OneDVtwoPlain & \TwoDVtwoPlain & \MDVthreePlain & \MDVVPlain\\
\cmidrule(){1-14}
3gp100& & \underline{6.2} & 820.6 & 181.8 & & 14.3 & \underline{14.0} & 16.5 & & 18.4 & \underline{16.7} & 430.6 & 79.0\\
3r150& & \underline{19.8} & 1\,737.9 & 65.7 & & 17.6 & 18.7 & \underline{17.1} & & 22.7 & \underline{18.9} & 147.8 & 45.4\\
4gp30& & \underline{1.5} & 150.3 & 45.0 & & \underline{0.7} & 1.2 & 1.1 & & \underline{1.4} & 1.4 & 116.9 & 17.5\\
4r80& & \underline{10.5} & 987.5 & 64.5 & & \underline{7.9} & 18.0 & 15.3 & & \underline{11.2} & 18.4 & 344.8 & 98.2\\
5gp12& & \underline{0.3} & 38.5 & 3.6 & & \underline{0.2} & 0.4 & 0.5 & & \underline{0.5} & 0.5 & 30.6 & 1.6\\
5r40& & \underline{16.9} & 425.9 & 34.3 & & \underline{2.3} & 7.2 & 6.3 & & \underline{4.6} & 8.6 & 386.9 & 35.3\\
\bottomrule
\end{tabular}
\end{table}

\begin{table}[ht] \centering
\scriptsize
\caption{Local search heuristics started from Greedy.}
\label{tab:greedy}
\begin{tabular}{@{} l @{\hspace{0.5em}}  c @{\hspace{2em}} *{7}{r @{\hspace{0.5em}}} r @{} c @{\hspace{2em}} *{7}{r @{\hspace{0.5em}}} r @{}}
\toprule
 &  & \multicolumn{8}{c}{Solution error, \%} &  & \multicolumn{8}{c}{Running times, ms}\\
\cmidrule(){3-10}
\cmidrule(){12-19}
Inst. &  & \TwooptPlain & \OneDVPlain & \TwoDVPlain & \MDVPlain & \OneDVtwoPlain & \TwoDVtwoPlain & \MDVthreePlain & \MDVVPlain &  & \TwooptPlain & \OneDVPlain & \TwoDVPlain & \MDVPlain & \OneDVtwoPlain & \TwoDVtwoPlain & \MDVthreePlain & \MDVVPlain\\
\cmidrule(){1-19}
3gp100& & 4.3 & 3.4 & 3.4 & 3.4 & 3.4 & 3.4 & \underline{3.3} & 3.4 & & \underline{0.04} & 0.04 & 0.04 & 0.04 & 0.05 & 0.05 & 0.36 & 0.09\\
3r150& & 16.7 & 1.2 & 1.2 & 1.2 & 1.2 & 1.2 & 0.8 & \underline{0.7} & & \underline{0.02} & 0.02 & 0.02 & 0.03 & 0.02 & 0.03 & 0.11 & 0.05\\
4gp30& & 4.5 & 3.7 & 3.6 & 3.6 & 3.6 & 3.6 & \underline{2.6} & 3.6 & & 0.04 & \underline{0.03} & 0.04 & 0.04 & 0.04 & 0.04 & 0.11 & 0.05\\
4r80& & 15.8 & 7.9 & 6.1 & 6.1 & 7.9 & 6.1 & 2.6 & \underline{1.5} & & \underline{0.01} & 0.02 & 0.02 & 0.02 & 0.02 & 0.02 & 0.21 & 0.08\\
5gp12& & 5.4 & 6.3 & 4.5 & 4.5 & 5.3 & 4.5 & \underline{1.8} & 4.5 & & \underline{0.01} & 0.01 & 0.01 & 0.01 & 0.01 & 0.01 & 0.03 & 0.01\\
5r40& & 18.5 & 19.8 & 13.5 & 13.5 & 15.0 & 13.5 & \underline{2.3} & 3.5 & & \underline{0.01} & 0.01 & 0.01 & 0.01 & 0.01 & 0.01 & 0.18 & 0.04\\
\bottomrule
\end{tabular}
\end{table}

\begin{table}[ht] \centering
\scriptsize
\caption{Chain metaheuristic started from Trivial, Greedy and ROM\@.  5 seconds given.  1 --- \TwooptPlain, 2 --- \OneDVPlain, 3 --- \TwoDVPlain, 4 --- \MDVPlain, 5 --- \OneDVtwoPlain, 6 --- \TwoDVtwoPlain, 7 --- \MDVthreePlain, 8 --- \MDVVPlain.}
\label{tab:chain5}
\begin{tabular}{@{} l@{\hspace{0.5em}} c @{\hspace{2em}} *{7}{r @{\hspace{0.5em}}} r @{} c @{\hspace{2em}} *{7}{r @{\hspace{0.5em}}} r @{} c @{\hspace{2em}} *{7}{r @{\hspace{0.5em}}} r @{}}
\toprule
 &  & \multicolumn{26}{c}{Solution error, \%}\\
\cmidrule(){3-28}
 &  & \multicolumn{8}{c}{Trivial} &  & \multicolumn{8}{c}{Greedy} &  & \multicolumn{8}{c}{ROM}\\
\cmidrule(){3-10}
\cmidrule(){12-19}
\cmidrule(){21-28}
Inst. &  & 1 & 2 & 3 & 4 & 5 & 6 & 7 & 8 &  & 1 & 2 & 3 & 4 & 5 & 6 & 7 & 8 &  & 1 & 2 & 3 & 4 & 5 & 6 & 7 & 8\\
\cmidrule(){1-28}
3gp100& & 15.3 & \underline{1.8} & \underline{1.8} & \underline{1.8} & 1.8 & 1.8 & 2.8 & 2.5 & & 5.3 & 1.7 & \underline{1.7} & \underline{1.7} & 1.8 & 1.8 & 2.9 & 2.3 & & 9.8 & \underline{1.9} & \underline{1.9} & \underline{1.9} & 1.9 & 1.9 & 2.6 & 2.3\\
3r150& & 77.7 & \underline{0.0} & \underline{0.0} & \underline{0.0} & \underline{0.0} & \underline{0.0} & 0.1 & \underline{0.0} & & 41.4 & \underline{0.0} & \underline{0.0} & \underline{0.0} & \underline{0.0} & \underline{0.0} & \underline{0.0} & \underline{0.0} & & 33.6 & \underline{0.0} & \underline{0.0} & \underline{0.0} & \underline{0.0} & \underline{0.0} & \underline{0.0} & \underline{0.0}\\
4gp30& & 7.0 & 1.6 & 1.1 & 1.1 & 1.8 & 1.1 & \underline{0.8} & 1.4 & & 6.3 & 1.9 & 0.8 & \underline{0.8} & 1.8 & 0.9 & \underline{0.8} & 1.4 & & 2.2 & 1.7 & 0.9 & 0.9 & 1.7 & 1.0 & \underline{0.8} & 1.4\\
4r80& & 55.0 & 4.4 & 1.9 & 1.9 & 4.1 & 2.3 & 0.4 & \underline{0.0} & & 41.6 & 4.6 & 1.6 & 1.6 & 4.5 & 1.8 & 0.8 & \underline{0.0} & & 57.0 & 4.3 & 2.0 & 2.0 & 4.3 & 2.0 & 0.9 & \underline{0.0}\\
\bottomrule
\end{tabular}
\end{table}

\begin{table}[ht] \centering
\scriptsize
\caption{Chain metaheuristic started from Trivial, Greedy and ROM\@.  10 seconds given.  1 --- \TwooptPlain, 2 --- \OneDVPlain, 3 --- \TwoDVPlain, 4 --- \MDVPlain, 5 --- \OneDVtwoPlain, 6 --- \TwoDVtwoPlain, 7 --- \MDVthreePlain, 8 --- \MDVVPlain.}
\label{tab:chain10}
\begin{tabular}{@{} l@{\hspace{0.5em}} c @{\hspace{2em}} *{7}{r @{\hspace{0.5em}}} r @{} c @{\hspace{2em}} *{7}{r @{\hspace{0.5em}}} r @{} c @{\hspace{2em}} *{7}{r @{\hspace{0.5em}}} r @{}}
\toprule
 &  & \multicolumn{26}{c}{Solution error, \%}\\
\cmidrule(){3-28}
 &  & \multicolumn{8}{c}{Trivial} &  & \multicolumn{8}{c}{Greedy} &  & \multicolumn{8}{c}{ROM}\\
\cmidrule(){3-10}
\cmidrule(){12-19}
\cmidrule(){21-28}
Inst. &  & 1 & 2 & 3 & 4 & 5 & 6 & 7 & 8 &  & 1 & 2 & 3 & 4 & 5 & 6 & 7 & 8 &  & 1 & 2 & 3 & 4 & 5 & 6 & 7 & 8\\
\cmidrule(){1-28}
3gp100& & 15.1 & \underline{1.6} & \underline{1.6} & \underline{1.6} & 1.7 & 1.7 & 2.3 & 2.2 & & 5.3 & 1.6 & 1.6 & \underline{1.6} & 1.6 & 1.6 & 2.5 & 2.1 & & 9.8 & \underline{1.6} & \underline{1.6} & \underline{1.6} & 1.8 & 1.7 & 2.2 & 2.1\\
3r150& & 75.3 & \underline{0.0} & \underline{0.0} & \underline{0.0} & \underline{0.0} & \underline{0.0} & \underline{0.0} & \underline{0.0} & & 41.4 & \underline{0.0} & \underline{0.0} & \underline{0.0} & \underline{0.0} & \underline{0.0} & \underline{0.0} & \underline{0.0} & & 33.6 & \underline{0.0} & \underline{0.0} & \underline{0.0} & \underline{0.0} & \underline{0.0} & \underline{0.0} & \underline{0.0}\\
4gp30& & 6.5 & 1.4 & 0.8 & 0.8 & 1.3 & 1.0 & \underline{0.7} & 1.3 & & 6.2 & 1.7 & 0.8 & 0.8 & 1.5 & 0.8 & \underline{0.7} & 1.1 & & 2.2 & 1.4 & 0.8 & 0.8 & 1.4 & 0.8 & \underline{0.7} & 1.2\\
4r80& & 52.1 & 3.9 & 1.1 & 1.0 & 3.6 & 1.1 & 0.1 & \underline{0.0} & & 41.4 & 3.9 & 1.0 & 1.0 & 4.3 & 1.1 & 0.4 & \underline{0.0} & & 55.0 & 4.0 & 1.1 & 1.1 & 3.4 & 1.3 & 0.4 & \underline{0.0}\\
\bottomrule
\end{tabular}
\end{table}

\begin{table}[ht] \centering
\scriptsize
\caption{Multichain metaheuristic started from Trivial, Greedy and ROM\@.  5 seconds given.  1 --- \TwooptPlain, 2 --- \OneDVPlain, 3 --- \TwoDVPlain, 4 --- \MDVPlain, 5 --- \OneDVtwoPlain, 6 --- \TwoDVtwoPlain, 7 --- \MDVthreePlain, 8 --- \MDVVPlain.}
\label{tab:multichain5}
\begin{tabular}{@{} l@{\hspace{0.5em}} c @{\hspace{2em}} *{7}{r @{\hspace{0.5em}}} r @{} c @{\hspace{2em}} *{7}{r @{\hspace{0.5em}}} r @{} c @{\hspace{2em}} *{7}{r @{\hspace{0.5em}}} r @{}}
\toprule
 &  & \multicolumn{26}{c}{Solution error, \%}\\
\cmidrule(){3-28}
 &  & \multicolumn{8}{c}{Trivial} &  & \multicolumn{8}{c}{Greedy} &  & \multicolumn{8}{c}{ROM}\\
\cmidrule(){3-10}
\cmidrule(){12-19}
\cmidrule(){21-28}
Inst. &  & 1 & 2 & 3 & 4 & 5 & 6 & 7 & 8 &  & 1 & 2 & 3 & 4 & 5 & 6 & 7 & 8 &  & 1 & 2 & 3 & 4 & 5 & 6 & 7 & 8\\
\cmidrule(){1-28}
3gp100& & 11.8 & \underline{1.1} & 1.2 & \underline{1.1} & 1.3 & 1.3 & 156.9 & 2.0 & & 5.3 & 1.2 & 1.2 & \underline{1.1} & 1.4 & 1.3 & 5.6 & 2.2 & & 9.7 & \underline{1.2} & \underline{1.2} & \underline{1.2} & 1.3 & 1.3 & 9.8 & 2.0\\
3r150& & 68.1 & \underline{0.0} & \underline{0.0} & \underline{0.0} & \underline{0.0} & \underline{0.0} & \underline{0.0} & \underline{0.0} & & 41.4 & \underline{0.0} & \underline{0.0} & \underline{0.0} & \underline{0.0} & \underline{0.0} & \underline{0.0} & \underline{0.0} & & 33.6 & \underline{0.0} & \underline{0.0} & \underline{0.0} & \underline{0.0} & \underline{0.0} & \underline{0.0} & \underline{0.0}\\
4gp30& & 3.0 & 0.9 & \underline{0.7} & \underline{0.7} & 0.8 & \underline{0.7} & \underline{0.7} & 1.1 & & 3.1 & 0.8 & \underline{0.7} & \underline{0.7} & 1.0 & \underline{0.7} & \underline{0.7} & 1.0 & & 2.1 & \underline{0.7} & \underline{0.7} & \underline{0.7} & \underline{0.7} & \underline{0.7} & 0.8 & 1.0\\
4r80& & 45.3 & 3.4 & 1.5 & 1.4 & 2.3 & 1.5 & 0.4 & \underline{0.0} & & 38.9 & 3.3 & 0.9 & 0.9 & 3.1 & 0.9 & 0.8 & \underline{0.0} & & 44.6 & 2.4 & 1.0 & 1.0 & 2.6 & 1.1 & 45.3 & \underline{0.0}\\
\bottomrule
\end{tabular}
\end{table}

\begin{table}[ht] \centering
\scriptsize
\caption{Multichain metaheuristic started from Trivial, Greedy and ROM\@.  10 seconds given.  1 --- \TwooptPlain, 2 --- \OneDVPlain, 3 --- \TwoDVPlain, 4 --- \MDVPlain, 5 --- \OneDVtwoPlain, 6 --- \TwoDVtwoPlain, 7 --- \MDVthreePlain, 8 --- \MDVVPlain.}
\label{tab:multichain10}
\begin{tabular}{@{} l@{\hspace{0.5em}} c @{\hspace{2em}} *{7}{r @{\hspace{0.5em}}} r @{} c @{\hspace{2em}} *{7}{r @{\hspace{0.5em}}} r @{} c @{\hspace{2em}} *{7}{r @{\hspace{0.5em}}} r @{}}
\toprule
 &  & \multicolumn{26}{c}{Solution error, \%}\\
\cmidrule(){3-28}
 &  & \multicolumn{8}{c}{Trivial} &  & \multicolumn{8}{c}{Greedy} &  & \multicolumn{8}{c}{ROM}\\
\cmidrule(){3-10}
\cmidrule(){12-19}
\cmidrule(){21-28}
Inst. &  & 1 & 2 & 3 & 4 & 5 & 6 & 7 & 8 &  & 1 & 2 & 3 & 4 & 5 & 6 & 7 & 8 &  & 1 & 2 & 3 & 4 & 5 & 6 & 7 & 8\\
\cmidrule(){1-28}
3gp100& & 11.2 & \underline{1.0} & \underline{1.0} & \underline{1.0} & 1.1 & 1.1 & 2.5 & 1.7 & & 5.3 & \underline{1.0} & \underline{1.0} & \underline{1.0} & 1.1 & 1.1 & 2.8 & 1.8 & & 9.7 & 1.0 & \underline{0.9} & 1.0 & 1.2 & 1.1 & 2.3 & 1.7\\
3r150& & 65.3 & \underline{0.0} & \underline{0.0} & \underline{0.0} & \underline{0.0} & \underline{0.0} & \underline{0.0} & \underline{0.0} & & 41.4 & \underline{0.0} & \underline{0.0} & \underline{0.0} & \underline{0.0} & \underline{0.0} & \underline{0.0} & \underline{0.0} & & 33.6 & \underline{0.0} & \underline{0.0} & \underline{0.0} & \underline{0.0} & \underline{0.0} & \underline{0.0} & \underline{0.0}\\
4gp30& & 2.9 & \underline{0.7} & \underline{0.7} & \underline{0.7} & \underline{0.7} & \underline{0.7} & \underline{0.7} & 1.0 & & 2.6 & 0.8 & \underline{0.7} & \underline{0.7} & \underline{0.7} & \underline{0.7} & \underline{0.7} & 0.9 & & 2.1 & \underline{0.7} & \underline{0.7} & \underline{0.7} & \underline{0.7} & \underline{0.7} & \underline{0.7} & 0.8\\
4r80& & 43.8 & 2.5 & 0.9 & 0.9 & 2.1 & 1.0 & 0.2 & \underline{0.0} & & 38.1 & 2.3 & 0.6 & 0.6 & 2.3 & 0.6 & 0.5 & \underline{0.0} & & 42.4 & 2.0 & 0.8 & 0.8 & 2.0 & 0.8 & 0.5 & \underline{0.0}\\
\bottomrule
\end{tabular}
\end{table}

\begin{table}[ht] \centering
\scriptsize
\caption{Heuristics comparison for the instances with independent weights.}
\label{tab:heuristics_independent}
%
\end{table}

\begin{table}[ht] \centering
\scriptsize
\caption{Heuristics comparison for the instances with decomposable weights.}
\label{tab:heuristics_decomposable}
%
\\
\bottomrule
\end{tabular}
\end{table}

\end{document}